\documentclass[twocolumn]{aastex701}
\usepackage{caption}
\usepackage{float}
\usepackage{multirow}
\usepackage{subcaption}

\DeclareCaptionLabelFormat{parenright}{#2)} 
\shorttitle{Sensitivity to Companions in WFC3/IR}
\shortauthors{Mehta et al.}

\begin{document}
\title{The Hubble Ultracool Multiplicity (HUM) Survey. I. Characterizing Sensitivity to Companions at Sub-Diffraction Limit Separations with HST WFC3/IR}

\author[0009-0008-6318-7630]{Kunal Mehta}
\affiliation{Department of Astronomy, University of Michigan,
Ann Arbor, MI 48104, USA}
\affiliation{Department of Physics, Applied Physics, and Astronomy, Rensselaer Polytechnic Institute,
Troy, NY 12180, USA}
\email{kunal9mehta@gmail.com}

\author[0000-0003-1863-4960]{Matthew De Furio}
\affiliation{Department of Astronomy, College of Natural Sciences, University of Texas at Austin,
Austin, TX 78712, USA}
\email{defurio@utexas.com}

\author[0000-0001-8170-7072]{Daniella Bardalez Gagliuffi}
\affiliation{Department of Physics \& Astronomy, Amherst College, Amherst, MA 01002, USA}
\email{dbardalezgagliuffi@amherst.edu}

\author[0000-0001-9823-1445]{Trent J. Dupuy}
\affiliation{Institute for Astronomy, University of Edinburgh, Royal Observatory, Edinburgh, EH9 3HJ, UK}
\email{tdupuy@gmail.com}

\author[0000-0002-2428-9932]{Cl\'{e}mence Fontanive}
\affiliation{Institute for Astronomy, University of Edinburgh, Royal Observatory, Edinburgh, EH9 3HJ, UK}
\email{clemence.fontanive@umontreal.ca}

\author[0000-0001-9811-568X]{Adam L. Kraus}
\affiliation{Department of Astronomy, College of Natural Sciences, University of Texas at Austin,
Austin, TX 78712, USA}
\email{alk@astro.as.utexas.edu}

\author[0000-0003-1227-3084]{Michael R. Meyer}
\affiliation{Department of Astronomy, University of Michigan,
Ann Arbor, MI 48104, USA}
\email{mrmeyer@umich.edu}

\author[0009-0007-7333-8925]{Matthew Cole}
\affiliation{Institute for Astronomy, University of Edinburgh, Royal Observatory, Edinburgh, EH9 3HJ, UK}
\email{m.cole@roe.ac.uk}

\author[0009-0005-9614-466X]{Fernanda Sophia Morais Laroca}
\affiliation{Department of Physics \& Astronomy, Amherst College, Amherst, MA 01002, USA}
\email{fmoraislaroca27@amherst.edu}
\begin{abstract}
We characterize the sensitivity of a double point-spread function (PSF) fitting algorithm---employing empirical, position-dependent PSF models---for detecting companions using the infrared channel of the Wide Field Camera 3 (WFC3/IR) on the Hubble Space Telescope (HST). The observed separation distribution of known brown dwarf (BD) binaries is potentially biased towards separations larger than the angular resolution limits of current techniques. Previous imaging analyses suffer from incompleteness at separations $<2\lambda/D$; our aim is to probe within this limit to identify previously missed companions. We evaluate the performance of our technique on artificial data across 8 WFC3/IR filters and a broad range of signal-to-noise ratios (S/N), determining our ability to accurately recover injected companions and identifying the region of parameter space where false positive fits are likely. Here, we demonstrate the capability of this technique to recover companions at sub-pixel separations on the WFC3/IR detector---below the diffraction limit in multiple filters. For F160W at a typical S/N of 75, we resolve companions separated by 0.8 pixels (104 mas, $0.759\lambda/D$) at 1.5 magnitudes contrast with $>90\%$ confidence. We achieve the closest angular resolution for any detection method with WFC3/IR imaging to date. Compared to previous BD multiplicity surveys with WFC3/IR, we achieve a 2.5$\times$ improvement in separation sensitivity at contrasts of 0-3 magnitudes in F127M. We have demonstrated that applying our improved technique to archival HST images of field BDs will thus probe down to separations of 1 au, in one of the largest high angular resolution surveys of such objects to date.
\end{abstract}


\section{Introduction} \label{sec:intro}

Multiplicity is a common outcome of the star and brown dwarf (BD) formation process and has two main mechanisms: turbulent fragmentation and disk fragmentation. Turbulent fragmentation can produce binary and higher order multiple systems across the full range of primary masses and orbital separations out to 10,000s au \citep{2023ASPC..534..275O}. Disk fragmentation requires a massive disk to become unstable and can produce companions at separations on the order of the size of the disk, $\sim$ 100 au \citep{Kratter2010ApJ...708.1585K}. Brown dwarfs represent the low mass end of the turbulent fragmentation process, and while they are known to host disks \citep[e.g.,][]{Martinez2022AJ....163...36M}, it is unknown whether those disks can become unstable to produce companions. Subsequent processes may alter the initial conditions of a given binary system. Companions residing within the disk of the primary can preferentially accrete material, building up its mass relative to the primary. Interactions between the binary and the gas within the protostellar envelope can cause companions to migrate and reduce their separations \citep{Baruteau2011ApJ...726...28B}. These processes will influence the final properties of the population of multiple systems. In multiplicity surveys of the Galactic field population, companions to BD primaries are mainly detected at separations $<$ 20 au \citep{2003ApJ...587..407C} and their separation distribution can be modeled by a log-normal with a peak at 3-6 au \citep{2006AJ....132..891R,10.1093/mnras/sty1682,2019BAAS...51c.285B}.

Most companions to BDs have been found through high resolution imaging either from ground-based adaptive optics or space-based diffraction-limited optical/NIR imaging \citep[e.g.,][]{2015ApJ...803..102D,2019ApJ...883..205B,10.1093/mnras/stad2870,2006AJ....132..891R, 2006ApJS..166..585B, 2003ApJ...587..407C, 2003AJ....126.1526B}. Most multiplicity surveys of BDs typically do not achieve sensitivity to companions at the diffraction limit \citep[e.g.,][]{2014AJ....148..129A, 10.1093/mnras/stad2870} and are incomplete near the observed peak of the companion separation distribution, a potential observational bias \citep{2015AJ....150..163B}. Additionally, these surveys suffer from small sample sizes resulting in companion frequency errors on the order of 10\% \citep{2019BAAS...51c.285B}. Therefore, a large multiplicity survey of BDs uniformly analyzed to achieve diffraction limited resolution would place strong constraints on BD companion population properties, i.e. frequency, orbital separations, and mass ratios.

In the HST Archival program AR-17561 (co-PIs: D. Bardalez Gagliuffi, M. De Furio), we set out to fully analyze all field BDs observed with the Wide Field Camera 3 Infrared instrument (WFC3/IR) on the Hubble Space Telescope (HST) with a point-spread function (PSF) fitting technique proven to resolve companions down to the diffraction limit \citep[e.g.][]{2019ApJ...886...95D}. The sample consists of nearly 180 BDs, many of which were observed for other science purposes without investigating multiplicity, making it one of the largest multiplicity survey of BDs compiled to date.

While our technique has been demonstrated on the Advanced Camera for Survey Wide Field Camera (ACS/WFC) on HST \citep{2019ApJ...886...95D, 2022ApJ...925..112D, 2022ApJ...941..161D} and NIRCam \citep{2023ApJ...948...92D,2023ApJ...947L..30C} and MIRI \citep{2025ApJ...990L..63D} on the James Webb Space Telescope (JWST), we must first characterize our ability to detect companions across a number of filters in WFC3/IR and across a range of signal-to-noise ratios (S/N) to enable this multiplicity survey. In this paper, we present the results of our analysis, demonstrate the sensitivity of this technique with WFC3/IR, and display example recoveries of known binaries at a range of separations and contrasts.

In Sec. \ref{sec:methods}, we describe the WFC3/IR detector, our double-PSF model, and our fitting algorithm. In Sec. \ref{sec:analysis}, we describe our analysis to define sensitivity across eight WFC3/IR filters. In Sec. \ref{sec:results}, we present results of the sensitivity of our technique to companions for WFC3/IR. In Sec. \ref{sec:application}, we apply the technique to real data and demonstrate the ability to recover known companions near the diffraction limit. In Sec. \ref{sec:discussion}, we discuss the limitations, implications, and future of our work. In Sec. \ref{sec:conclusion}, we summarize our results.

\section{Methods} \label{sec:methods}
\subsection{WFC3/IR Binary Example Data} \label{subsec:wfc3}
The ultimate goal of our analysis is to define the limits of our technique on WFC3/IR images across filters useful for our science interests. These filters (F098M, F105W, F110W, F125W, F127M, F140W, F153M, and F160W) cover the wavelength range of 800-1700 nm. The IR channel has a $136''\times123''$ field of view with reference pixel plate scale of $0.128243''$/pixel \citep{2010wfc..rept....8K} and an average of $0.13''$/pixel \citep{2024wfci.book...17M}.

In Sec. \ref{sec:application}, we consider WFC3/IR images of four known binaries with varied separations and contrasts as test cases. WISE J033605.05-014350.4 was observed during GO-12970 (PI: Michael C. Cushing) in F125W on 2013-02-04 at 10:00:04 UT with an exposure time of 602.94 seconds. WISE J014656.66+423410.0 has an observation taken during SNAP-12873 (PI: Beth Biller) in F127M on 2013-06-24 at 20:07:40 UT with an exposure time of 349.23 seconds. CFBDSIR J145829+101343 has an observation taken during GO-12504 (PI: Michael C. Liu) in F125W on 2012-08-01 at 10:09:45 UT with an exposure time of 138.38 seconds. And finally, WISEPA J045853.89+643452.9 was observed during GO-12504 (PI: Michael C. Liu) in F160W on 2012-02-27 at 20:46:36 UT with an exposure time of 73.74 seconds. We used HST *\_flt.fits data products which are bias corrected, dark subtracted, and flat fielded exposures reduced with the calwf3 pipeline \citep{2024wfcd.book....6P}. The HST data presented in this article were obtained from the Mikulski Archive for Space Telescopes (MAST) at the Space Telescope Science Institute. The specific observations analyzed can be accessed via \dataset[doi: 10.17909/nn1m-yq16]{https://doi.org/10.17909/nn1m-yq16}.

\subsection{The Double-PSF Fitting Model} \label{subsec:model}
We utilize the double-PSF fitting technique described in \cite{2023ApJ...948...92D} and \cite{2023ApJ...947L..30C} which implements the Python module PyMultiNest \citep{2014A&A...564A.125B} built on the Nested Sampling Monte Carlo framework Multinest \citep{10.1111/j.1365-2966.2009.14548.x}. A summary of this algorithm is provided below.

Before fitting, we first subtract the mean background derived from an annulus around the source with an inner radius of 12 pixels and an outer radius of 15 pixels. For the examples in Sec. \ref{sec:application}, we inspected the annulus by eye to ensure no background contaminants were present. When applying this technique to our archival sample, we will implement automated filtering for background contaminants in the annuli. Since our data for this analysis was synthetic, filtering for contaminants was not necessary. We then identify the center of the target in image coordinates (generally the brightest pixel) and pass to the algorithm.

We use a grid of position-dependent, empirical PSF models \citep{2000PASP..112.1360A, 2016wfc..rept...12A}, each of which are 4$\times$ super-sampled and designed for the specific filter. For each pixel in our model PSF, a bicubic interpolation of the four closest empirical PSFs is performed. Then, a linear interpolation is used to combine these values based on the distance to the desired detector position for our model PSF. The double-PSF model is defined by six parameters: $x$ and $y$ position of the primary, the combined flux of the primary and companion, the separation in pixels, the position angle (PA) of the companion, and the contrast in magnitudes of the companion relative to the primary. 

We define priors for our parameters: $-2 \leq x_{\rm pri} \leq 2$ pixels, $-2 \leq y_{\rm pri} \leq 2$ pixels, $0 \leq \text{flux} \leq 10^6 $ electrons/second, $0 \leq \theta \leq 360$ degrees, $0.01 \leq s \leq 7$ pixels, and $0 \leq \text{contrast} \leq 6$ magnitudes. The origin $(x,y = 0,0)$ corresponds to the center of the target passed to the algorithm. The flux prior is uniform in log space while the rest are uniform in linear space. The algorithm returns the best-fit values to the six parameters, as well as the best-fit likelihood statistic which we define as the chi-squared statistic:
\begin{equation}
    \chi_{\nu}^2 = \frac{1}{\nu -1} \sum_{i=1}^{N}{\frac{(data_i - model_i)^2}{\sigma_i^2}}
    \label{eq:chisq}
\end{equation}
The algorithm will always return a best-fit binary solution to the data, so we must separately determine if that best-fit solution is statistically significant. We describe the ability to detect companions as a function of filter on WFC3/IR and S/N value in Sec. \ref{sec:analysis}. 

\setcounter{table}{1}
\begin{deluxetable*}{|c|cccccccc|}[!t]
\tablecaption{Vega magnitudes corresponding to each filter and S/N\label{tab:SNRvega}}
\tablehead{
\colhead{S/N} & \colhead{F098M} & \colhead{F105W} & \colhead{F110W} & 
\colhead{F125W} & \colhead{F127M} & \colhead{F140W} & \colhead{F153M} & \colhead{F160W}
}
\startdata
15  & 22.07  & 22.60  & 23.00   & 22.30  & 20.60  & 22.30  & 20.12  & 21.65 \\ 
\hline
35  & 21.10  & 21.65  & 22.10   & 21.35  & 19.65  & 21.40  & 19.20  & 20.70 \\
\hline
75  & 20.23  & 20.93  & 21.37  & 20.47  & 18.95  & 20.51  & 18.30  & 19.82 \\
\hline
150 & 19.39  & 20.06  & 20.50   & 19.62  & 18.09 & 19.66  & 17.45  & 18.97 \\
\hline
300 & 18.45  & 19.09  & 19.53   & 18.68  & 17.11  & 18.72  & 16.52  & 18.03 
\enddata
\tablecomments{We assume an exposure time of 400 seconds which is representative of imaging in our sample. Synthetic singles and binary systems were created for each magnitude and the S/N was calculated using Equation~\ref{eq:SNR}.}
\end{deluxetable*}
\subsection{Synthetic Image Creation}\label{subsec:image creation}
To define the sensitivity of our algorithm using WFC3/IR, we must characterize two features: the false positive probability of the binary fit and its ability to recover true companion parameters. To do this, we first created synthetic single sources and binary sources of various separations and contrasts utilizing the empirical PSF models for WFC3/IR \citep{2016wfc..rept...12A}. We make models for eight filters across S/N values of 15, 35, 75, 150, and 300 that are evenly spaced across the entire detector to sample the change in PSF shape as a function of position. Furthermore, the x and y coordinates of each centroid are randomly offset by a value between 0 and 1 pixels to model realistic observations and thoroughly sample sub-pixel positions. 

Each single is created within a 31 x 31 pixel postage stamp. Sources of noise shown, including photon noise, read noise, dark current, and sky background are also included as shown in table \ref{table:noise_values}.
\setcounter{table}{0}
\begin{deluxetable}{ccc}[H]
\tablecolumns{3}
\tablecaption{Values of different noise sources added to each pixel. 
\label{table:noise_values}}
\tablehead{
\colhead{Noise} & \colhead{Sampled Distribution} & \colhead{Value}}
\startdata
Photon Noise & Poisson & $\lambda =$ pixel value \\
Dark Current & Normal & 0.049 $\pm$ 0.007 e$^-$/s \\
Read Noise & Normal & 0.03 $\pm$ 0.01 e$^-$/s \\
Sky Noise & Normal & 0.56 $\pm$ 0.14 e$^-$/s 
\enddata
\end{deluxetable}
\vspace{-7mm}
Dark current \citep{2017wfc..rept....4S} and read noise \citep{2008wfc..rept...25H} are incorporated with values specific to WFC3/IR, while sky noise is derived from background averaged values of BD images representative of our sample taken in F160W during GO-17466 (PI: Clemence Fontanive) with an exposure time of 278 seconds. This value is conservatively representative of sky noise for our sample across all filters given the variation in exposure time. The ability to recover close companions is largely dependent on photon noise of the primary while for wider companions, recovery ability is limited by sky noise with modest contributions by dark current and read noise.

For each filter, 1089 singles (best fits in Fig. \ref{fig:fp_heatmap}) were created for Vega magnitudes corresponding to different S/N values, shown in table \ref{tab:SNRvega}. Since the postage stamp for each single is 31 x 31, we create an array of $33 \times 33 = 1089$ singles to sample the entire detector evenly. The magnitudes were calculated assuming an exposure time of 400 seconds. We assume this exposure time for all synthetic data since it is representative exposure time of the imaging in our sample that we aim to fit with our technique. S/N around each single was calculated within a 15 x 15 pixel postage stamp centered around the PSF:
\begin{equation}
    \text{S/N} = \frac{\sum_{i=1}^{N} (data_i - \overline{bkgd})}{\sqrt{\sum_{i=1}^{N} \sigma_i^2}}
    \label{eq:SNR}
\end{equation}

where $\sigma_i$ is defined as: 
\begin{equation}
    \text{$\sigma^{2}_i$} = \sigma_{Source}^2 + \sigma_{DC}^2 + \sigma_{RN}^2 + \sigma_{\overline{bkgd}}^2
    \label{eq:sigma}
\end{equation}
The same magnitudes apply to the synthetic binary systems since their magnitudes are defined as the combined flux of the primary and companion.

We create binaries using the same empirical PSF models, with the companion placed at a given position angle and separation relative to the primary and scaled to the flux ratio relative to the primary. To evenly sample the separation and contrast parameter spaces, each companion is created with separation, contrast, and position angle randomly sampled from uniform distributions. We generated more than 10,000 companions over separations of 0.01-4 pixels, position angles of 0-360 degrees, and contrasts of 0-6 magnitudes for each filter and S/N combination, see Fig. \ref{fig:comp_heatmap}. 
 \section{Analysis}\label{sec:analysis}
To determine the abilities of our algorithm on WFC3/IR data, we must define both the completeness rate and the false positive rate. A false positive analysis defines where in separation and contrast parameter space the algorithm is likely to converge when no companion is present. A completeness analysis tests the ability of the algorithm to accurately recover the known parameters of true synthetic binaries. 
\subsection{False Positive Analysis}\label{subsec:fpanalysis}
The purpose of the false positive analysis is to identify the region of parameter space in which the double-PSF code converges when applied to a known single source, thereby quantifying the likelihood that a given fit is consistent with a fit to a single PSF based on the recovered separation and contrast values. We create a unique contrast curve for every combination of filter and S/N, totaling 40.
\begin{figure}[h] 
    \centering
    \includegraphics[width=\columnwidth]{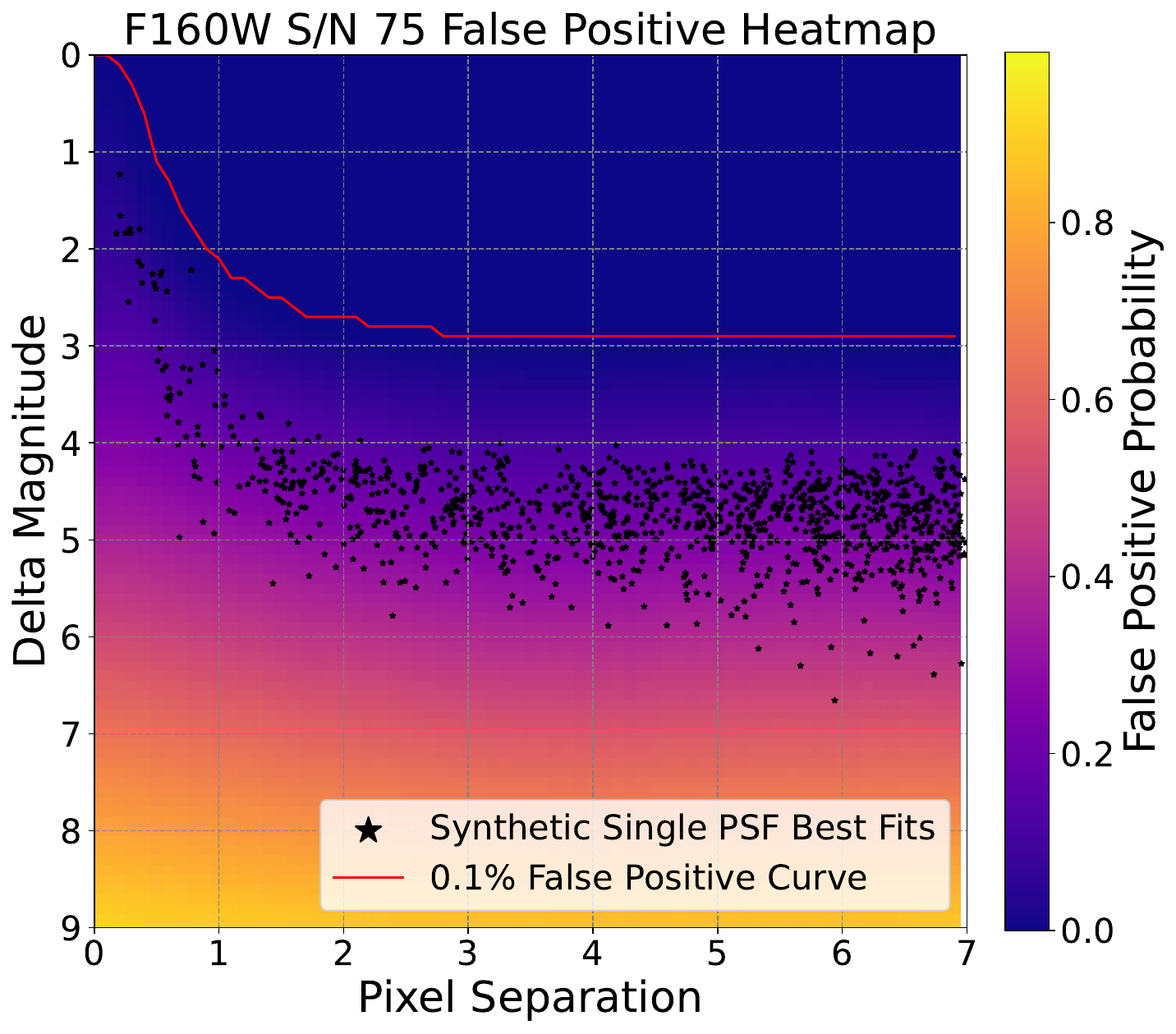} 
    \caption{False positive probability map in separation and contrast space of synthetic data in F160W with a S/N of 75. Overplotted is the 0.1\% false positive probability line and the black stars represent the best-fit binary parameters from the double-PSF fitting algorithm to 1089 artificial single PSFs. Since all 1089 fits fall below the 0.1\% probability line, fits above the line are inconsistent with a single PSF. The line appears higher than the black points because fits to singles can have individual samples of lower contrast than the best fit solution that have a false positive probability $>0.1\%$.}
    \label{fig:fp_heatmap}
\end{figure}
The method to characterize false positive occurrence follows the method from \cite{2022ApJ...925..112D}, and is summarized below. First, we run our double-PSF fitting algorithm on each of the 1089 synthetic singles. The fit of each single is the product of thousands of individual samples that the Multinest optimizer takes to arrive at a set of best-fit parameters. Therefore, each sample has an associated separation, contrast, and $\chi^2$ value. We convert the $\chi^2$ for each sample into a $\chi^2$ probability, $P(\chi^2)$. We then create bins of dimension 0.1 pixels separation by 0.1 magnitudes contrast and assign the value of each bin as the greatest $P(\chi^2)$ for all samples that fall within that bin. We normalize the distribution to sum to 1 across all bins. This results in a false positive probability distribution for one single. We then create a false positive probability distribution for each of the 1089 singles for a given S/N and filter, then sum all distributions and normalize the resulting distribution to 1. Finally, we normalize along each column of the same value in separation space to assign a cumulative false positive percentage to each bin. This value is defined by first assuming that the bin with the highest contrast allowed by the prior will have a false positive probability of 1, the lowest contrast will have a probability of zero, and each bin in the column is defined as the percent of false positive fits that will fall at lower contrasts, see Fig. \ref{fig:fp_heatmap}. 

\subsection{Completeness Analysis}\label{subsec:completeanalysis}
We evaluate our ability to recover true companions injected at a variety of separations and contrasts (as described in Sec. \ref{subsec:image creation}) by comparing recovered parameters to true parameters. This defines the likelihood in a region of parameter space that a given fit has accurate recovered companion parameters, with the definition of accurate given below. We create a unique contrast curve for every combination of filter and S/N, totaling 40. The method to characterize completeness follows from \cite{2019ApJ...886...95D}, and we summarize it below. 
\begin{figure}[H] 
    \centering
    \includegraphics[width=\columnwidth]{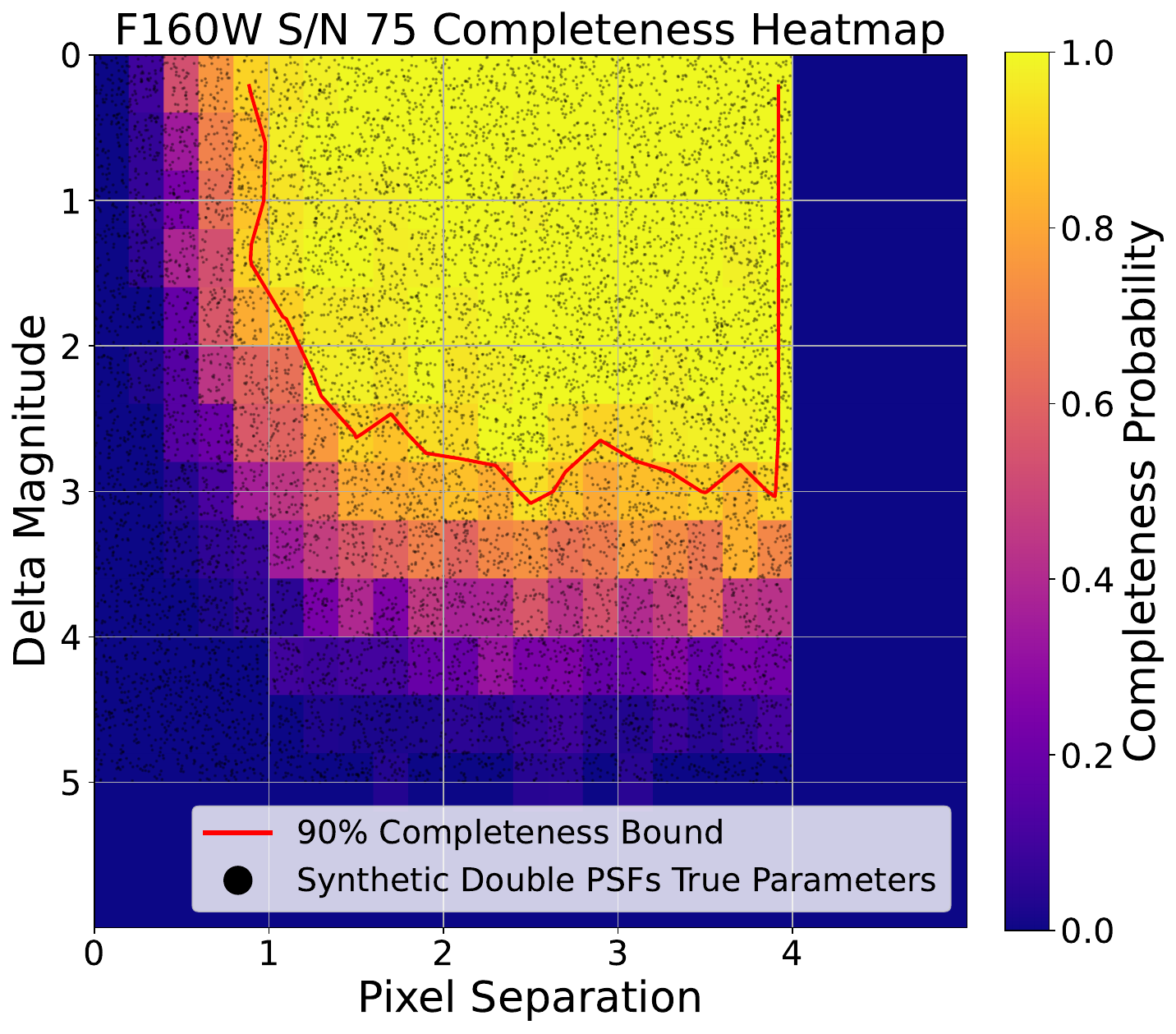} 
    \caption{Completeness probability map in separation and contrast of synthetic data in F160W with a S/N of 75. Overplotted is the 90\% completeness contour and the black stars represent the true synthetic companion parameters which were used to bin the data. The best-fit double-PSF recovered parameters were used to make the heatmap.}
    \label{fig:comp_heatmap}
\end{figure}
\vspace{-2mm}
We first apply our double-PSF fitting algorithm to more than 10,000 synthetic binary systems. We then define a companion as accurately recovered if the best-fit companion parameters are within a radius of 0.2 pixels and a contrast of 0.4 magnitudes of the true companion parameters. To set these bounds, we analyzed $\delta s = |s_t-s_r|$ and $\delta c = |c_t-c_r|$ where $s_t$ is the true separation, $s_r$ is the best-fit separation, $c_t$ is the true contrast, and $c_r$ is the best-fit contrast. For best fits along a set contrast range and inspecting decreasing $s_t$ bins, the distributions of $\delta s$ and $\delta c$ were concentrated around 0 with the $90^{\text{th}}$ percentile around 0.2 pixels and 0.4 magnitudes until closer separations where the distributions shifted to becoming uniform---indicative of a non-detection. The same occurred when isolating separation ranges and increasing bins in contrast. The $90^{\text{th}}$ percentile bounds were then chosen to define successful companion recovery.

To determine whether the recovered companion is within 0.2 pixels radius of the true companion, we use the following formula which incorporates position angle as well:
\begin{equation}
    \text{d} = \sqrt{{s_r}^2+{s_t}^2-2s_rs_t\cos{(\theta_r-\theta_t)}}
    \label{eq:polar distance}
\end{equation}
where $\theta_t$ is the true position angle, and $\theta_r$ is the recovered position angle. 

An issue with the fitting algorithm arises at low contrasts ($< 1$mag) where the recovered position angle can be offset from the true position angle by 180 degrees because the similar brightness primary and companion can be flipped in their designation as the primary object when calculating true parameters. We apply a correction for these cases that places the recovered companion at the position angle of the true companion. Then, completeness is calculated as above. 
To produce completeness maps (see Fig. \ref{fig:comp_heatmap}) we bin the separation and contrast space into bins of 0.4 magnitudes contrast by 0.2 pixels separation across the sampled range. We then define the value of each bin as the fraction of recovered companions. 
\section{Results}\label{sec:results}
\subsection{False Positive Curves}\label{subsec:fpcurves}
\begin{figure*}[h]
\gridline{\fig{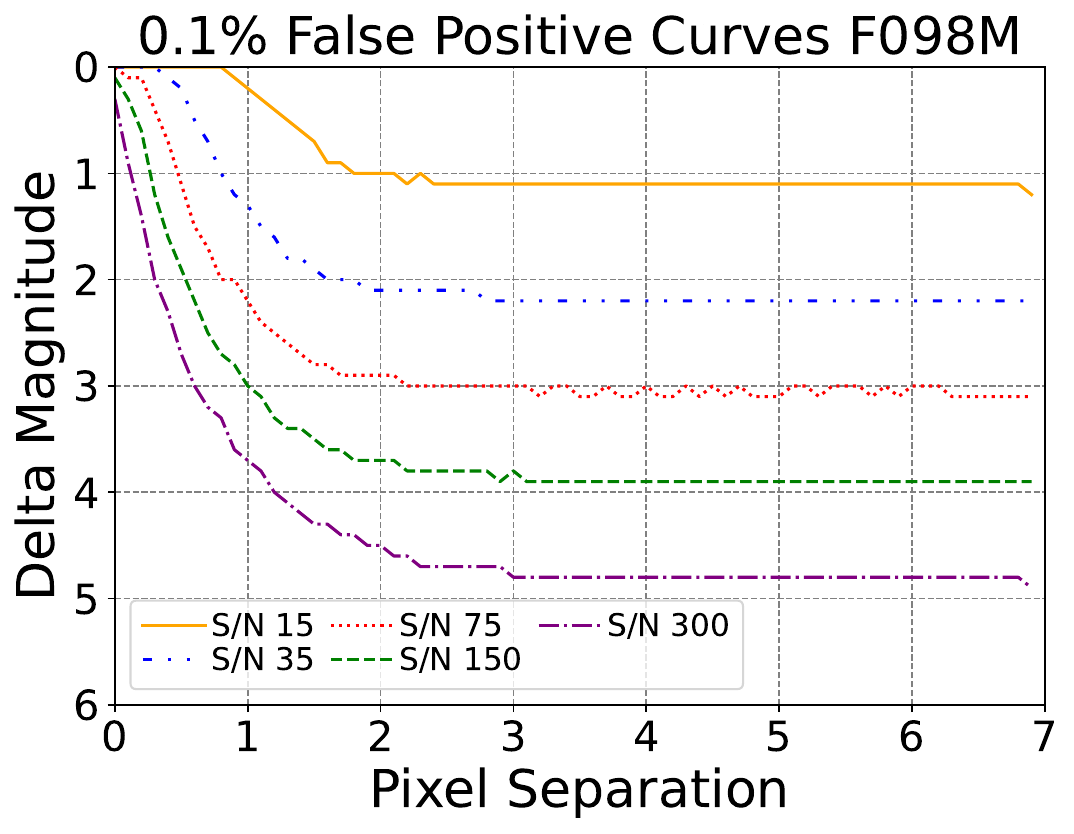}{0.48\textwidth}{\normalsize a) F098M}
          \fig{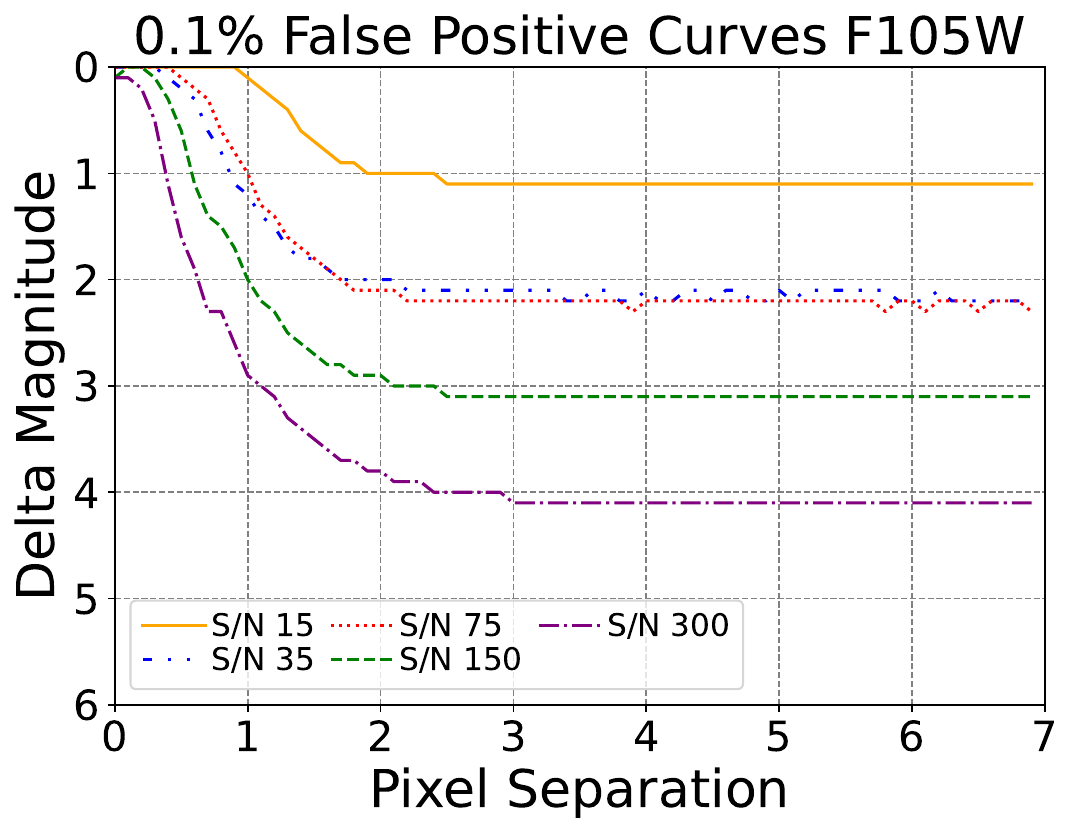}{0.48\textwidth}{\normalsize b) F105W}}
\vspace{-0.18cm}
\gridline{\fig{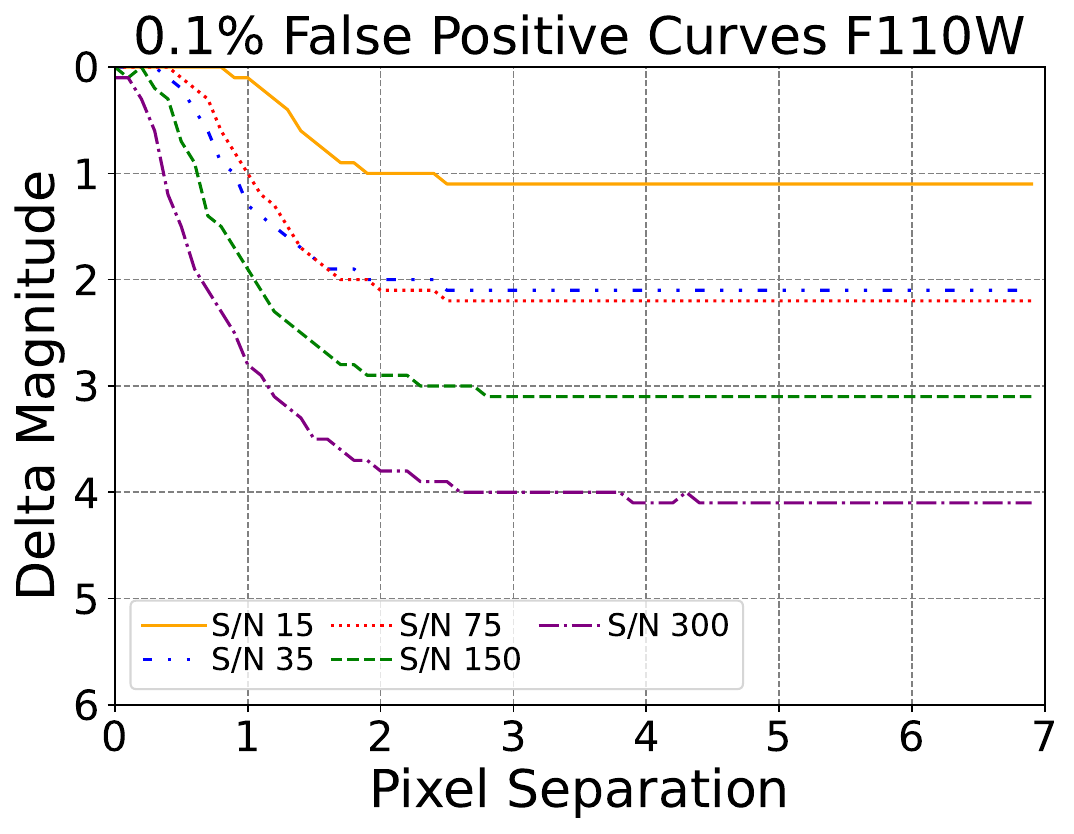}{0.48\textwidth}{\normalsize c) F110W}
          \fig{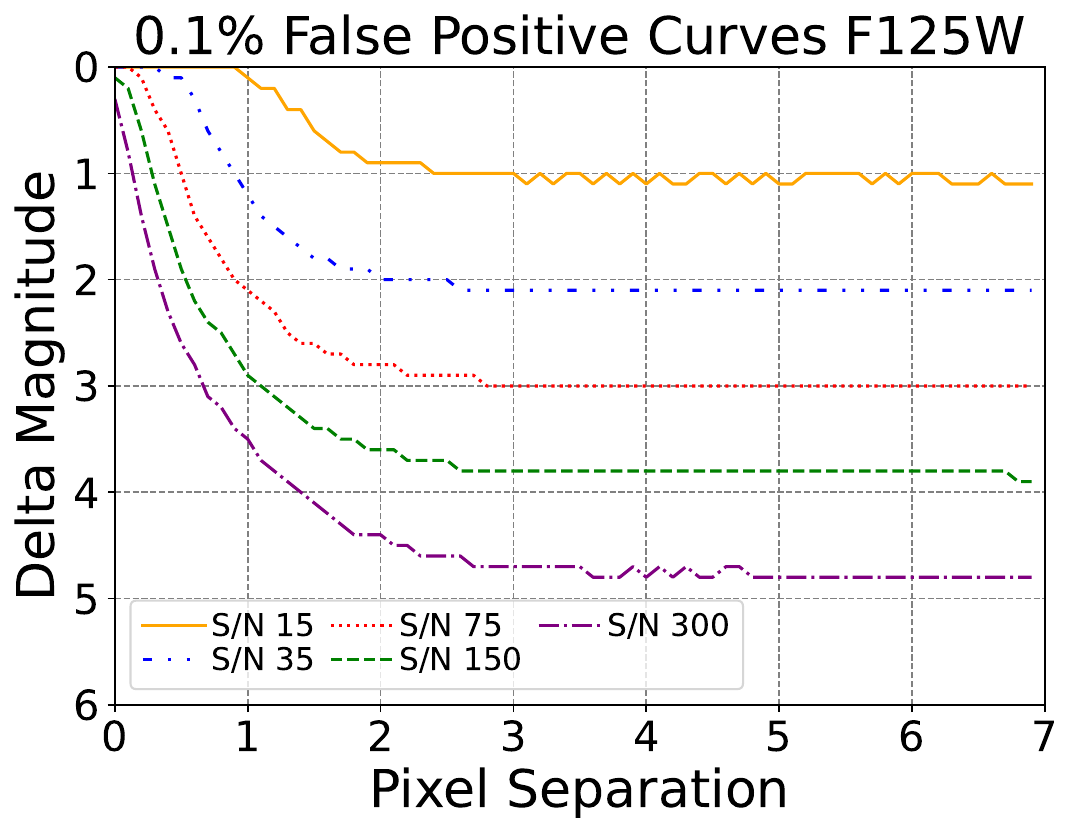}{0.48\textwidth}{\normalsize d) F125W}}
\vspace{-0.18cm}
\gridline{\fig{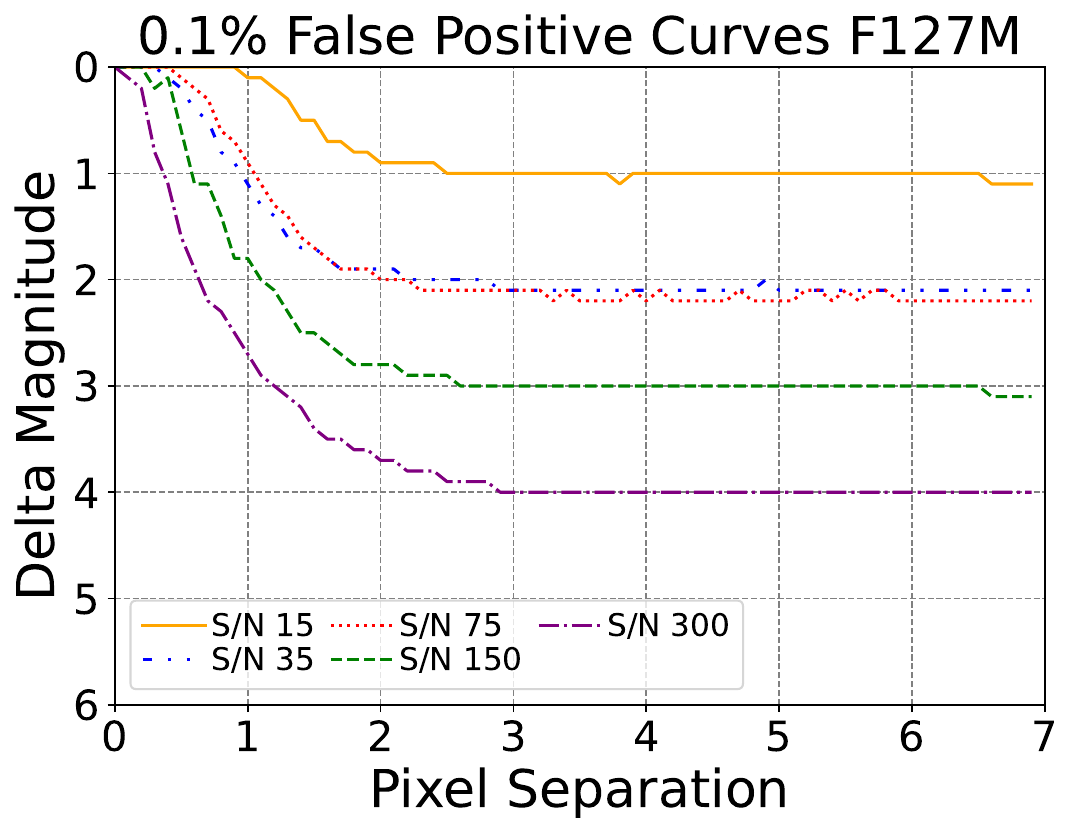}{0.48\textwidth}{\normalsize e) F127M}
          \fig{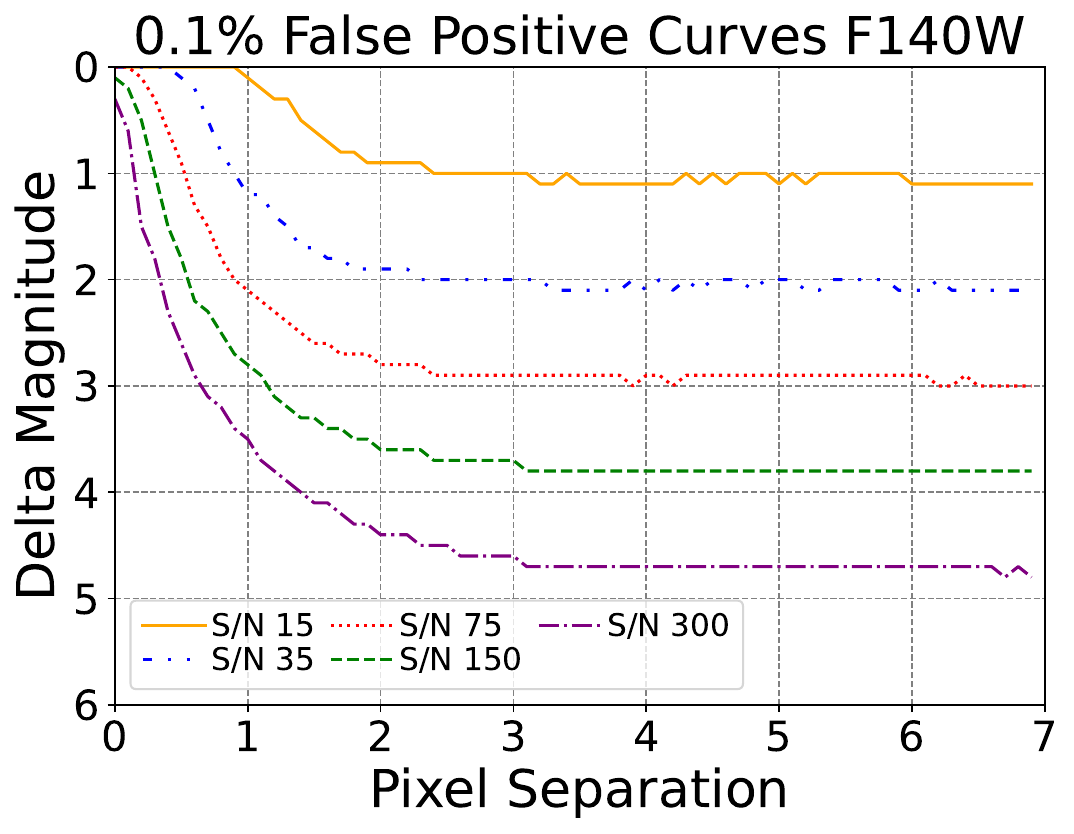}{0.48\textwidth}{\normalsize f) F140W}}
\phantomcaption
\end{figure*}
\begin{figure*}[t!]\ContinuedFloat
\gridline{\fig{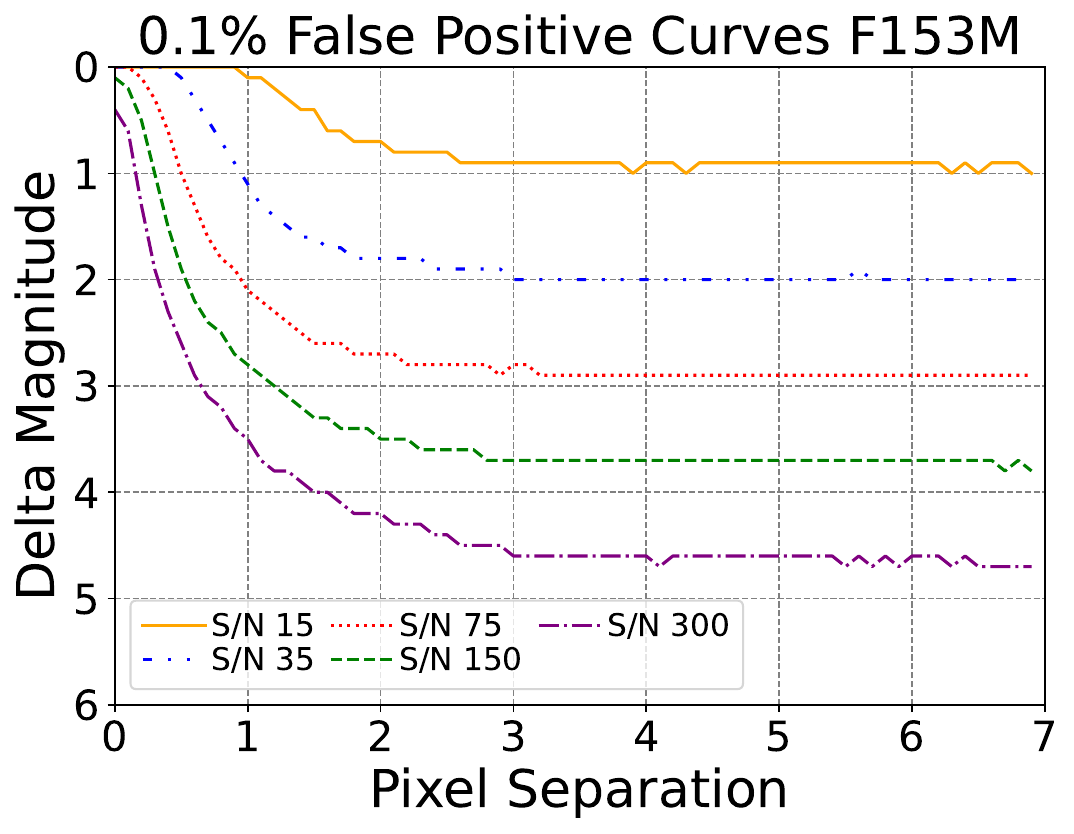}{0.48\textwidth}{\normalsize g) F153M}
          \fig{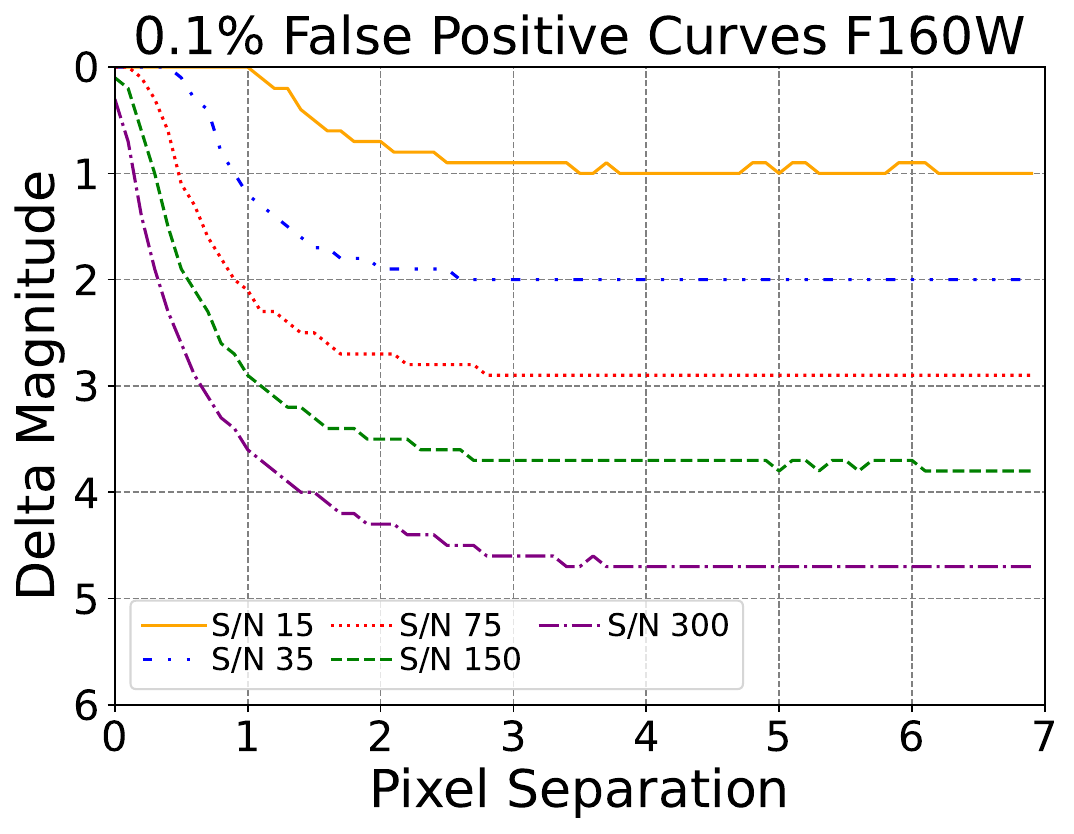}{0.48\textwidth}{\normalsize h) F160W}}
\caption{40 False-positive probability curves plotted by filter. As S/N increases, the area of the parameter space where detections are 99.9\% likely to not be false increases as well.}
\label{fig:fp_curves}
\end{figure*}
\renewcommand{\thefigure}{5}
\begin{figure*}[h]
\gridline{\fig{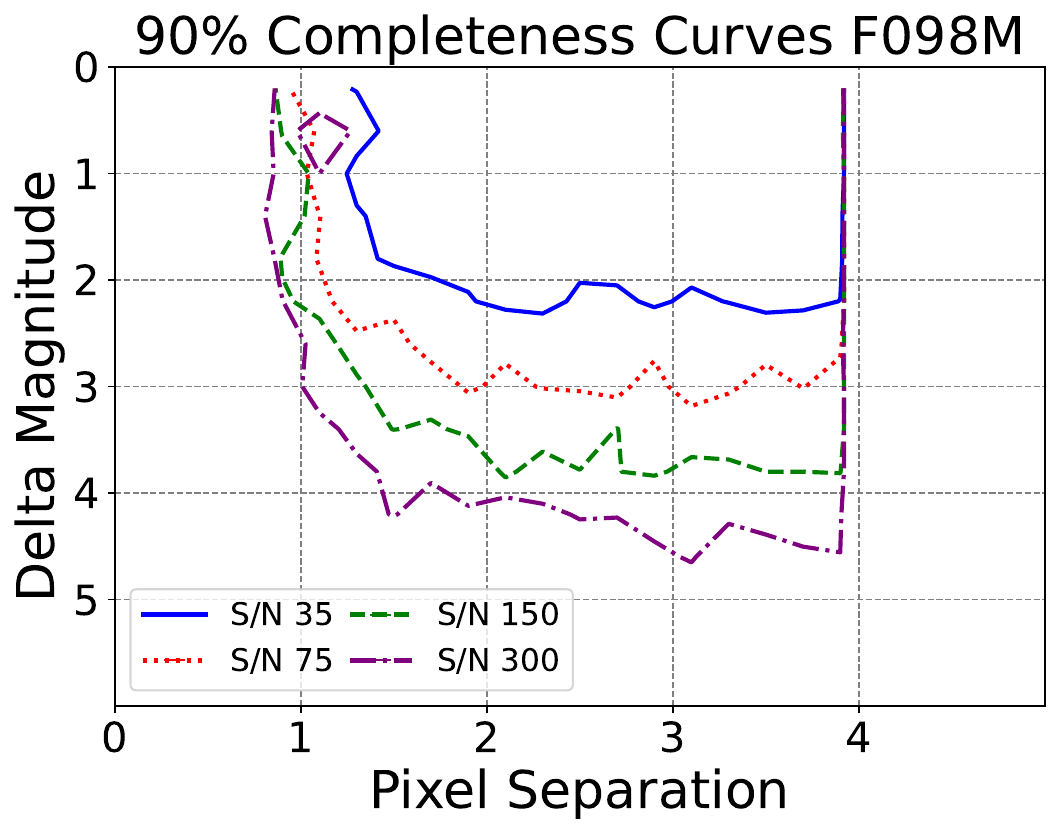}{0.47\textwidth}{\normalsize a) F098M}
          \fig{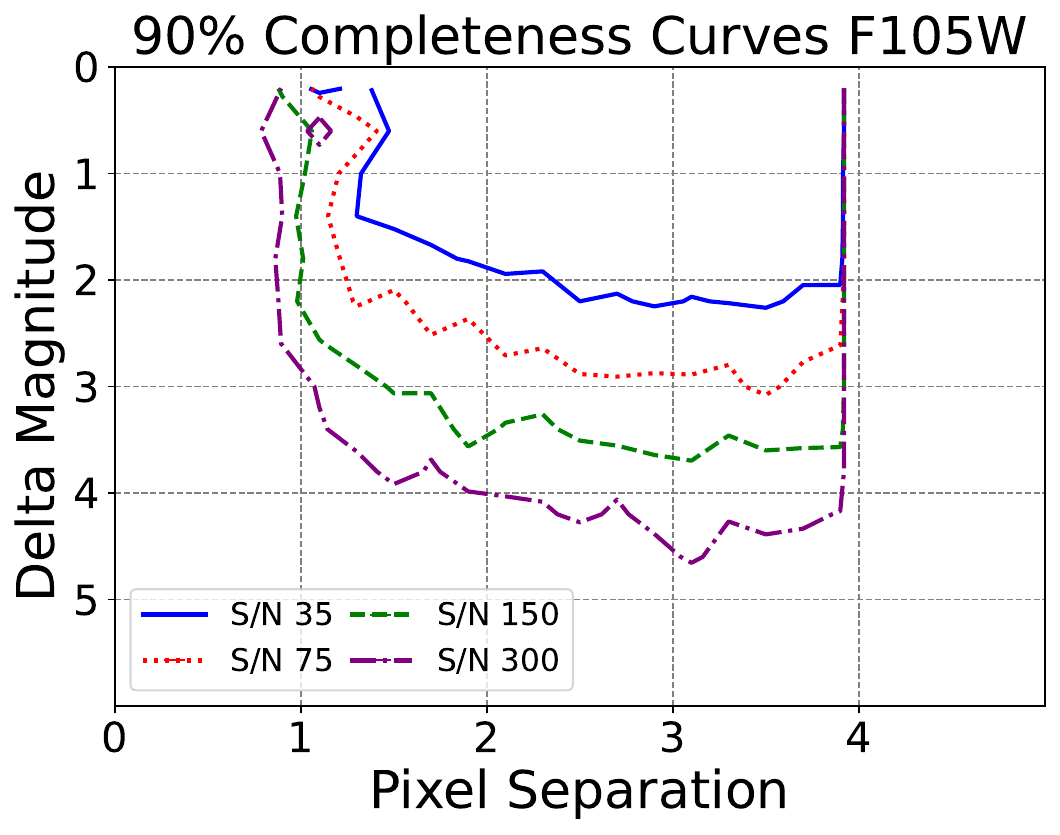}{0.47\textwidth}{\normalsize b) F105W}}
\vspace{-0.18cm}
\gridline{\fig{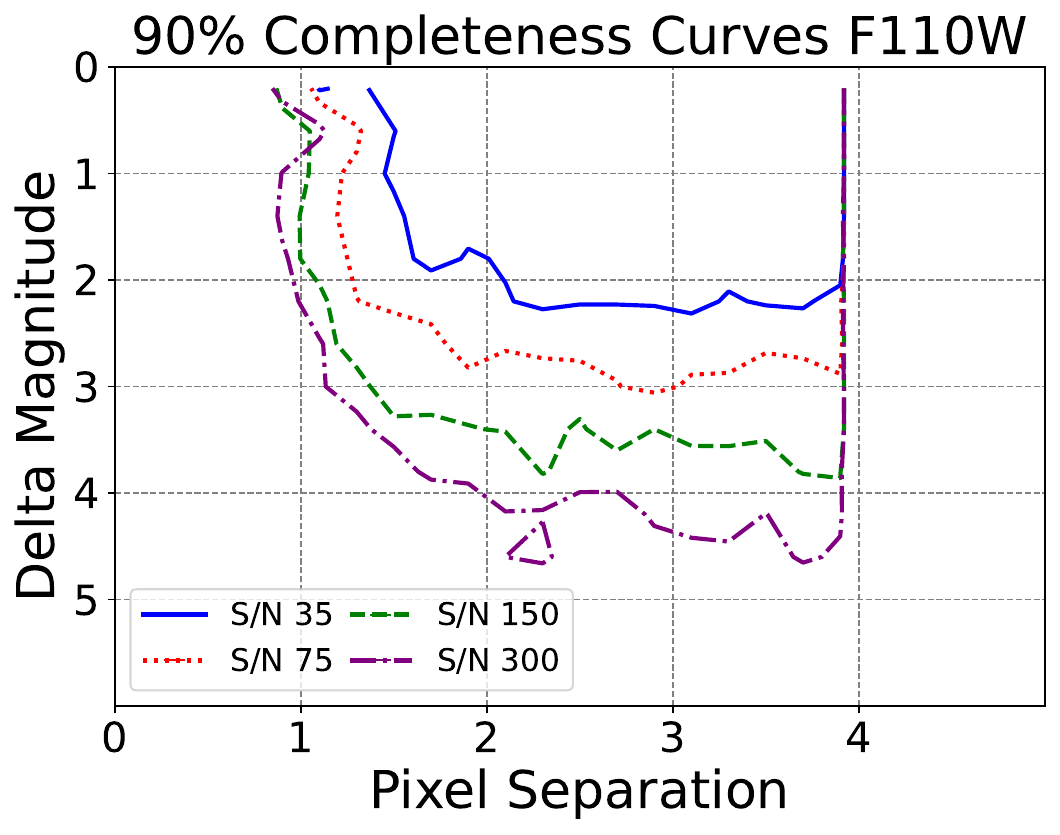}{0.47\textwidth}{\normalsize c) F110W}
          \fig{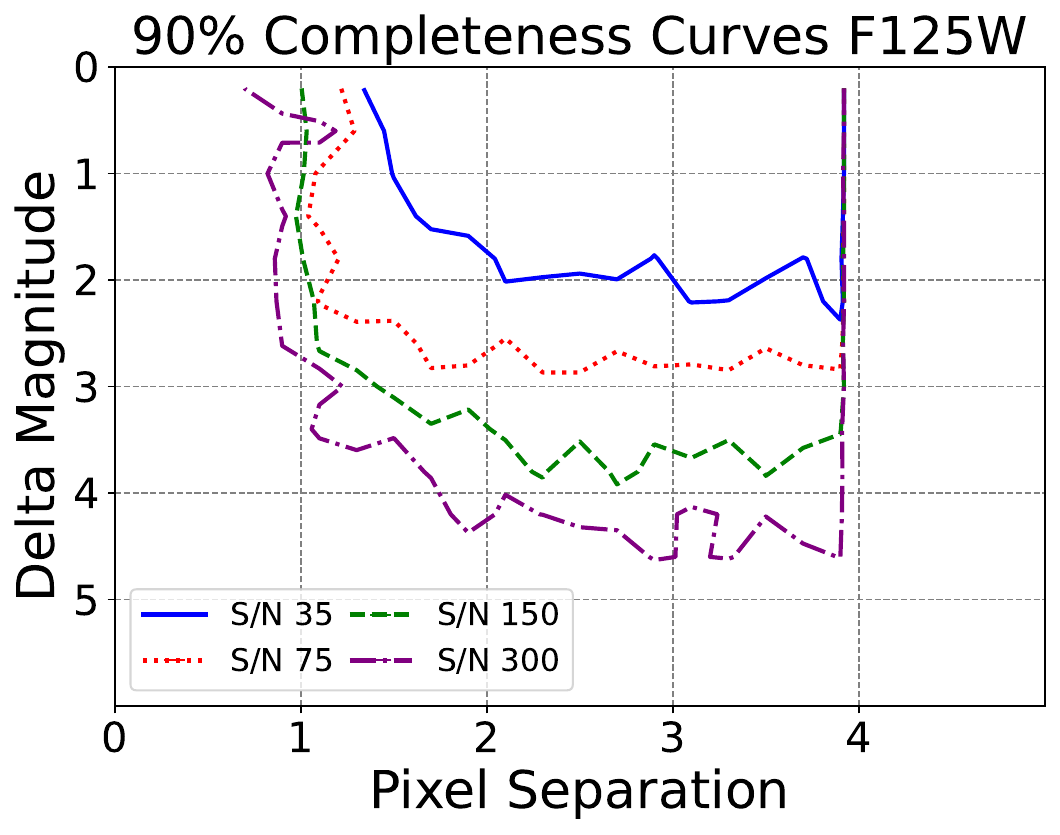}{0.47\textwidth}{\normalsize d) F125W}}
\vspace{-0.18cm}
\gridline{\fig{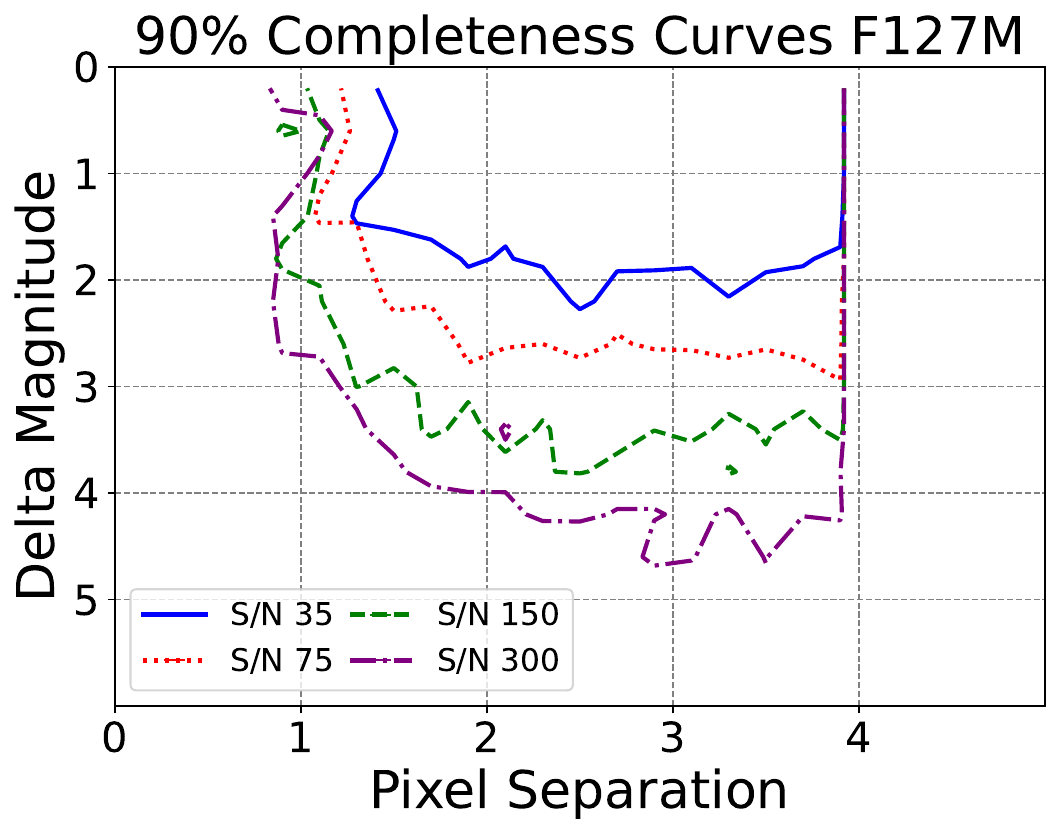}{0.47\textwidth}{\normalsize e) F127M}
          \fig{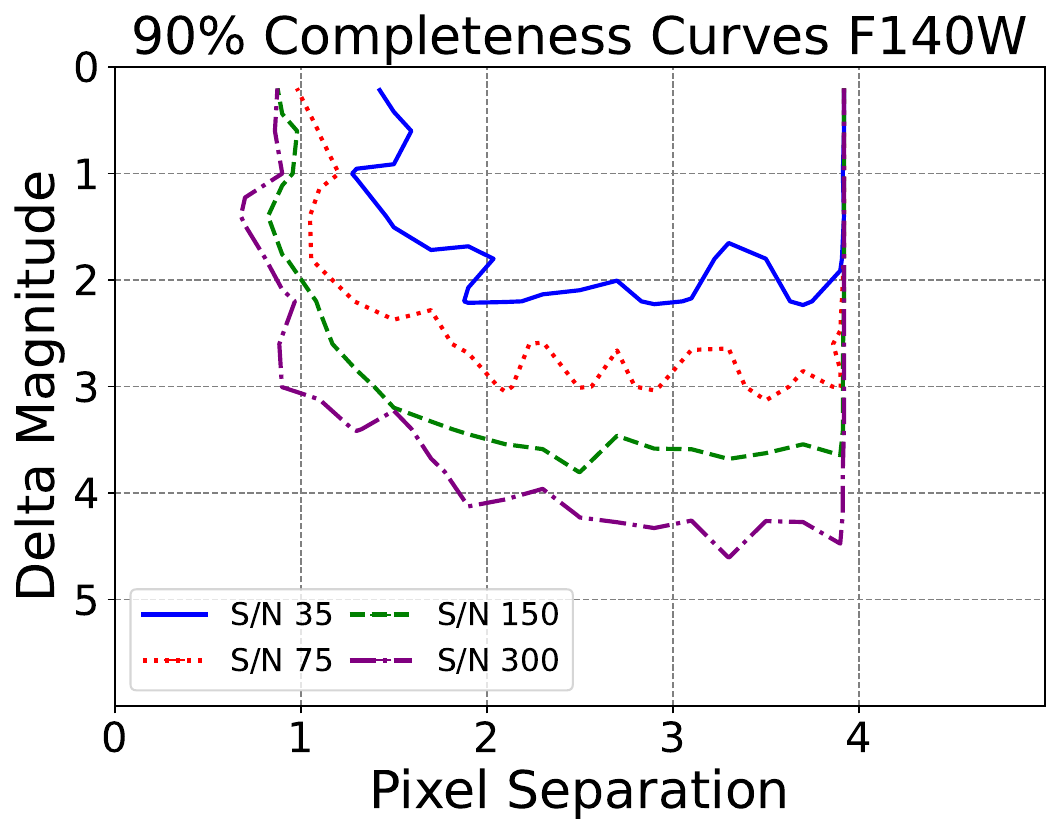}{0.47\textwidth}{\normalsize f) F140W}}
\phantomcaption
\end{figure*}
\begin{figure*}[t!]\ContinuedFloat
\gridline{\fig{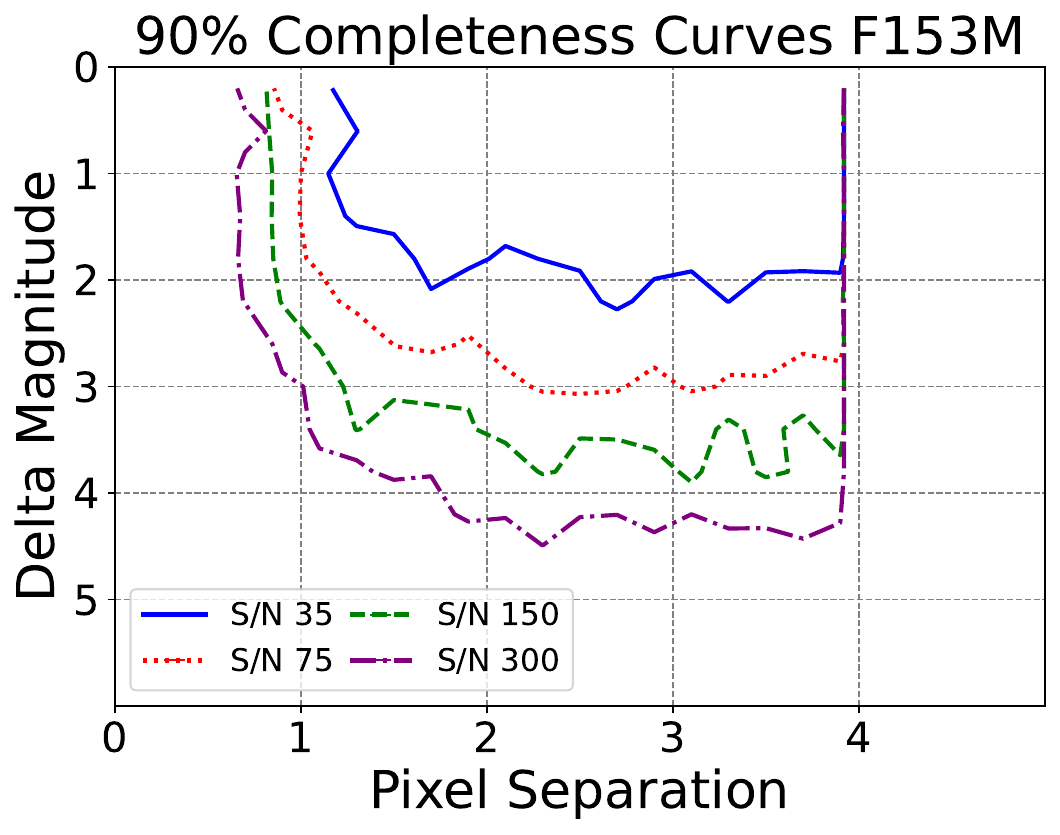}{0.47\textwidth}{\normalsize g) F153M}
          \fig{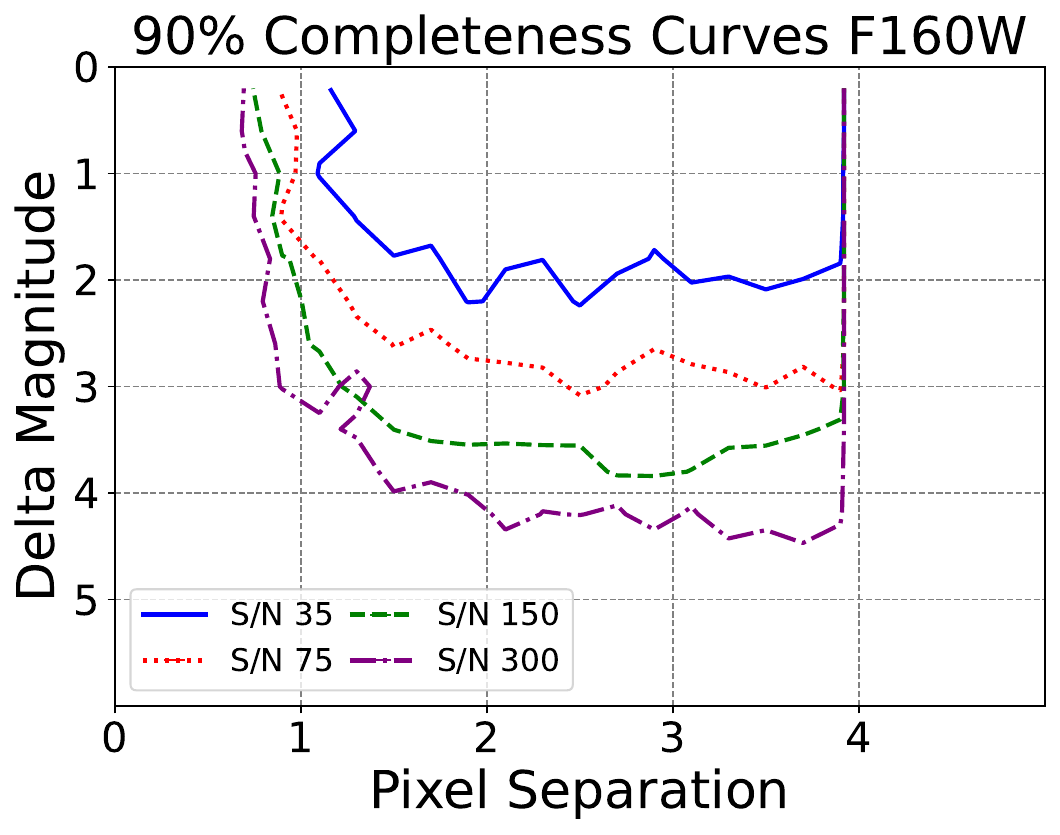}{0.47\textwidth}{\normalsize h) F160W}}
\caption{40 completeness probability curves plotted by filter. As S/N increases, we are able to accurately recover companion parameters with higher contrasts and closer separations.}
\label{fig:comp_curves}
\end{figure*}
We display 0.1\% false positive probability curves in Fig. \ref{fig:fp_curves}, showing the boundary below which 99.9\% of the probability of the false positive fits lies and above which a double-PSF fit to a single star will likely converge only 0.1\% of the time, as shown in Fig. \ref{fig:fp_heatmap}. Therefore, the code is unlikely to converge above this line when no companion is present. However, we must also determine how effectively we can recover true companions.

We display 40 distinct curves, 5 for each filter covering S/N values of 15, 35, 75, 150, 300. For real objects that have a S/N in between these values, a unique curve can be created by interpolating in S/N space. The false positive heatmap FITS files used to generate each curve (e.g. Fig \ref{fig:fp_heatmap}) are available in a .zip package for download with this publication.

\subsection{Completeness Curves}\label{subsec:completecurves}
In Fig. \ref{fig:comp_contours_15}, we show completeness curves of varying confidences for F160W with a S/N value of 15. In Fig. \ref{fig:comp_curves}, we display 90\% completeness probability curves for all filters and the remaining S/N values. Each curve denotes the region in separation and contrast space above which companion parameters are recovered accurately $\geq90\%$ of the time.

\renewcommand{\thefigure}{4}
\begin{figure}[H] 
    \centering
    \includegraphics[width=\columnwidth]{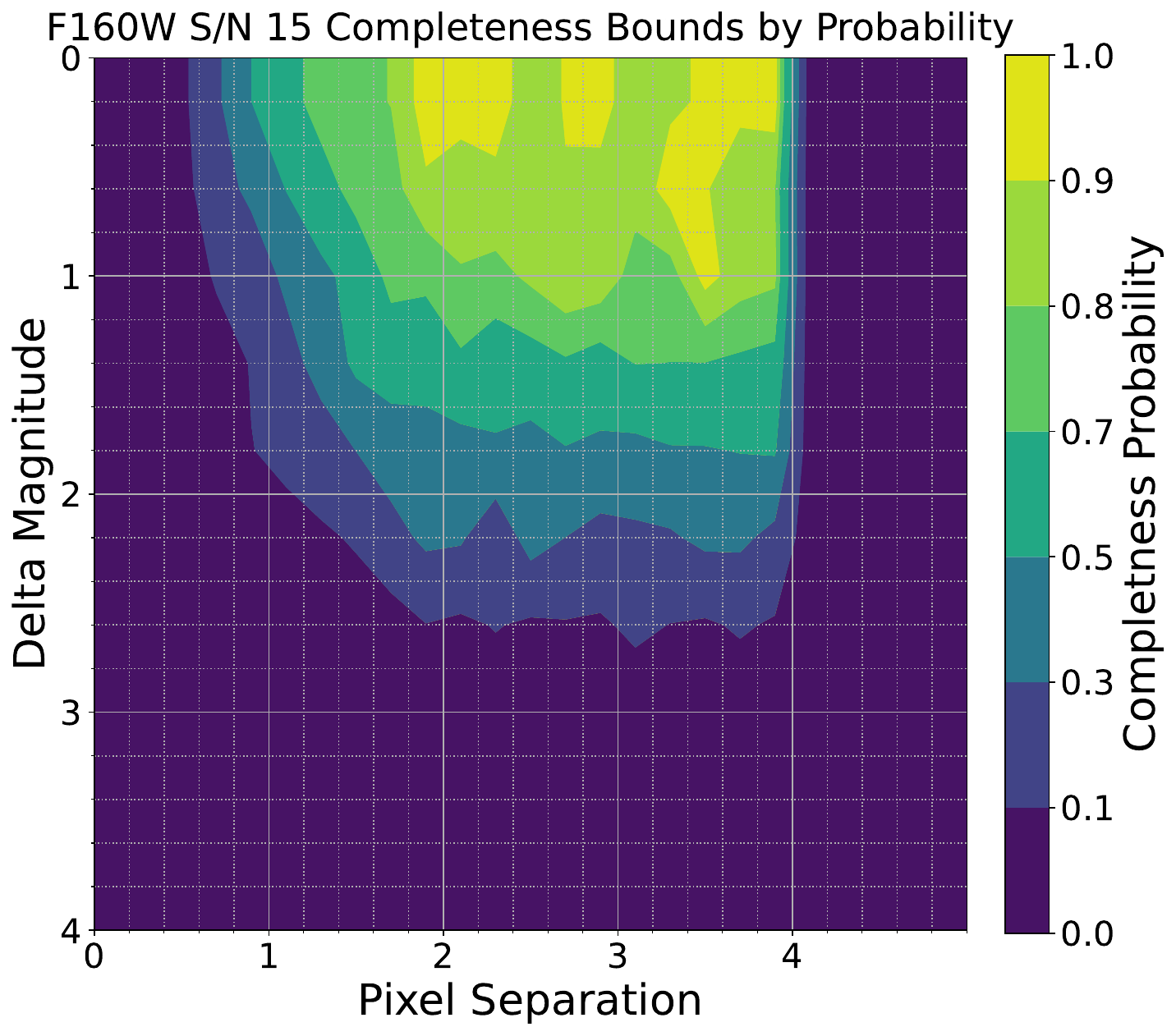} 
    \caption{Completeness probability contour plot in separation and contrast space of F160W with a S/N of 15. Since the completeness probability $> 90\%$ parameter space is fragmented, we would recover a true companion of this confidence less frequently than if the region was continuous. However, this does not mean recovery is impossible.}
    \label{fig:comp_contours_15}
\end{figure}
We show 35 distinct curves, 4 for each filter covering S/N values of 35, 75, 150, 300. For real objects that have a S/N value in between any of these, a unique curve can be created by interpolating in S/N space.

We also created heatmaps for S/N=15 in each filter, however the $>90\%$ completeness bounds across all filters are discontinuous and our technique's ability to recover companions to such faint primaries is less robust. In these cases, a lower completeness probability bound provides a smoother and more sensitive curve as shown in Fig \ref{fig:comp_contours_15}. 

The roughness and discontinuities present for some curves are the result of noise in confident recovery on the order of 1-3 binaries, since we created heatmaps that on average have 35-40 binaries per bin. We cut off sampling of the separation space at 4 pixels as the curves generally hit the background limit and our main science case is targeting close companions.

The completeness heatmap FITS files used to generate each curve (e.g. Fig \ref{fig:comp_heatmap}) are available in a .zip package for download with this publication.

\section{Application}\label{sec:application}
To validate the conclusions derived from our analysis of artificial sources, we apply our technique to four known BD binaries at varying separations and contrasts and display our ability to recover companions. These binaries were chosen from the The UltracoolSheet \citep[at http://bit.ly/UltracoolSheet;][]{best_w_m_j_2024_10573247}. We fit both double and single PSF models, then analyze the corresponding false positive and completeness probabilities associated with the best-fit parameters. We also present medians and uncertainties for the posterior distributions of separation and contrast, whereas the best-fit parameters are the fit from the maximum likelihood estimate.

\subsection{WISE J033605.05-014350.4}
WISE J033605.05-014350.4 \citep{2023ApJ...947L..30C} hereafter WISE 0336, is a Y+Y type binary recently discovered with JWST using the same double-PSF fitting technique as this paper. From \cite{2023ApJ...947L..30C}, imaging with JWST NIRcam places the separation at $89.8^{+3.8}_{-4.1}$ mas and places the contrast in F150W (closest filter to F127M) at $2.82^{+0.19}_{-0.11}$ magnitudes. This corresponds to a separation in WFC3/IR of $0.69^{+0.03}_{-0.03}$ pixels. We chose this system to display limits of our technique with a very close separation and relatively high contrast companion. 

Applying both our technique and a single-PSF fit, we obtain results shown in Fig. \ref{fig:wise0336_fits}. The posterior medians with large $1\sigma$ errors in PA ($90^{\circ}$), separation (1.6 pixels), and contrast (1.4 mags) from the double-PSF fit can be seen in table \ref{table:recovered_params} which support a very likely non-detection. The S/N in F125W for this object calculated using eq. \ref{eq:SNR} is $S/N \approx 4$ which is below the range characterized in the survey. We can compare with the F125W and $S/N=15$ completeness probability heatmap (see Fig. \ref{fig:wise0336_plot}), as the parameters falling within the bounds for a higher S/N value is a necessary condition if they are to do the same for a lower value. Given that the best fit separation (1.96 pixels) and contrast (2.00 magnitudes) result in a completeness probability of $46.4\%$ and a false positive probability of $9.4\%$, we cannot be confident that this detection is not a false positive, much less be confident in the accurate recovery of companion parameters given $S/N \approx 4$. Therefore, this binary at 0.84$\lambda/D$ represents a set of parameters that this technique is unable to recover, particularly in the close separation and low S/N regime. 
\renewcommand{\thefigure}{6}
\begin{figure*}[t!]
\centering
\begin{subfigure}[t]{0.47\textwidth}
    \includegraphics[width=\textwidth]{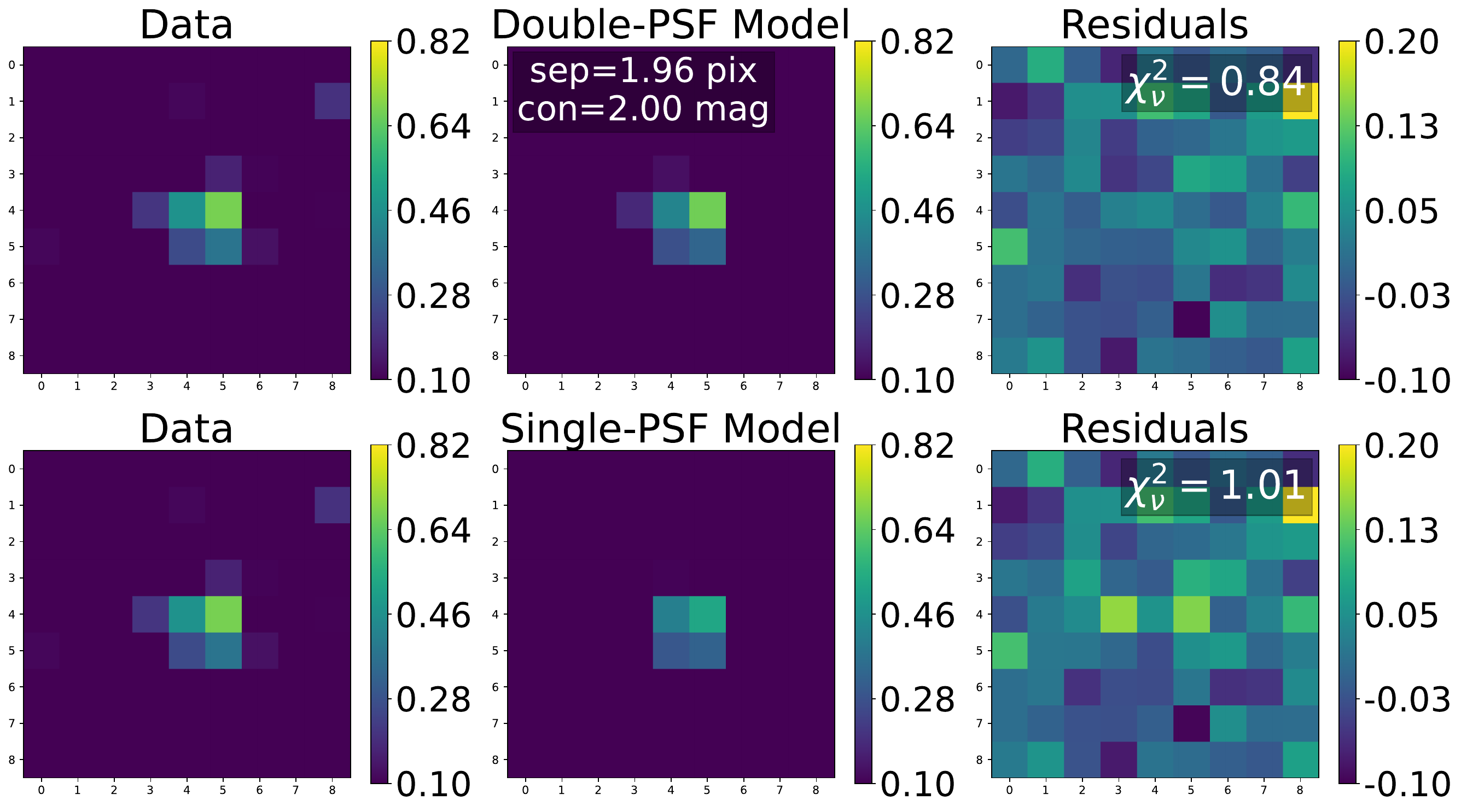}
    \caption{WISE 0336}
    \label{fig:wise0336_fits}
\end{subfigure}
\hfill
\begin{subfigure}[t]{0.47\textwidth}
    \includegraphics[width=\textwidth]{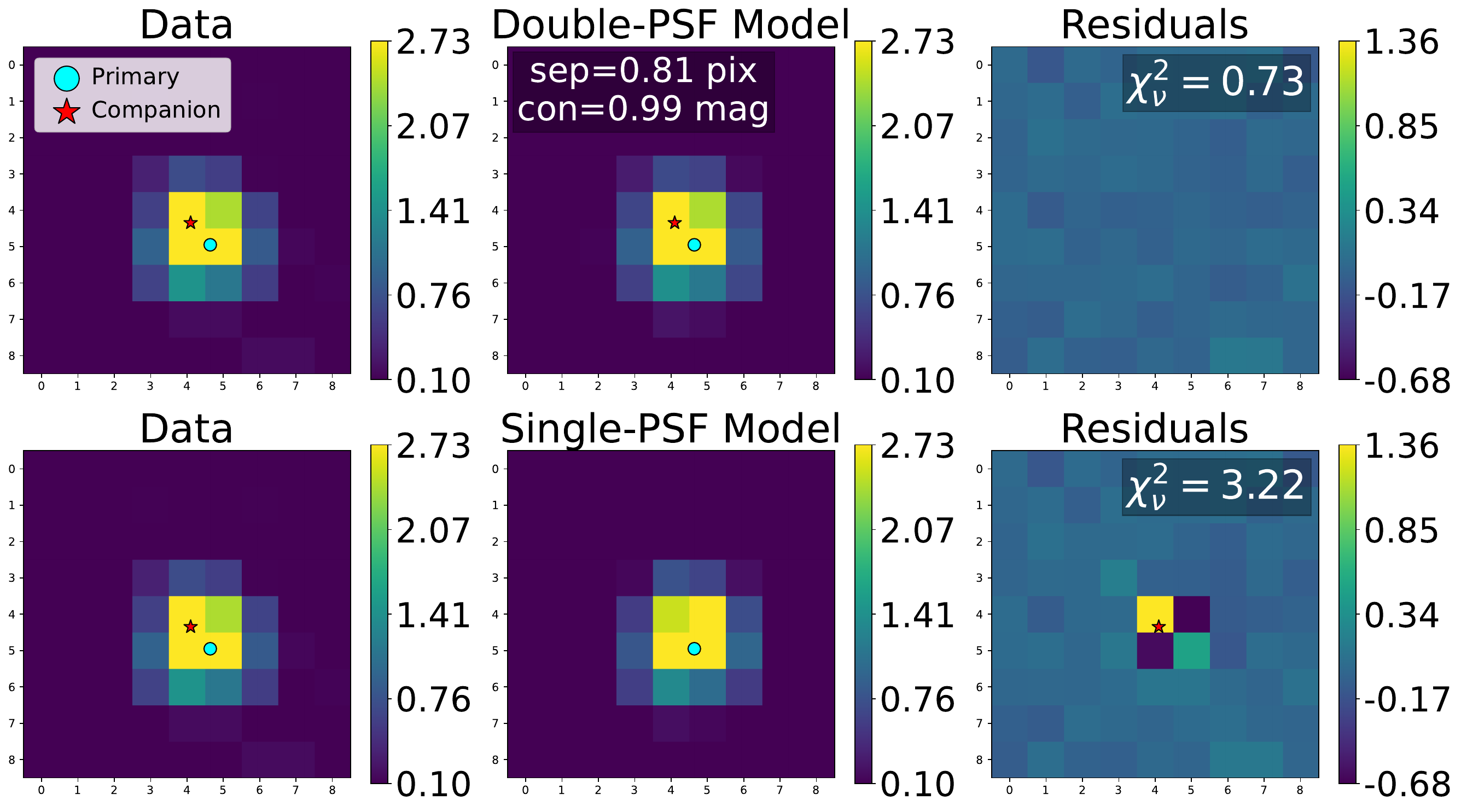}
    \caption{WISE 0146}
    \label{fig:wise0146_fits}
\end{subfigure}
\begin{subfigure}[t]{0.47\textwidth}
    \includegraphics[width=\textwidth]{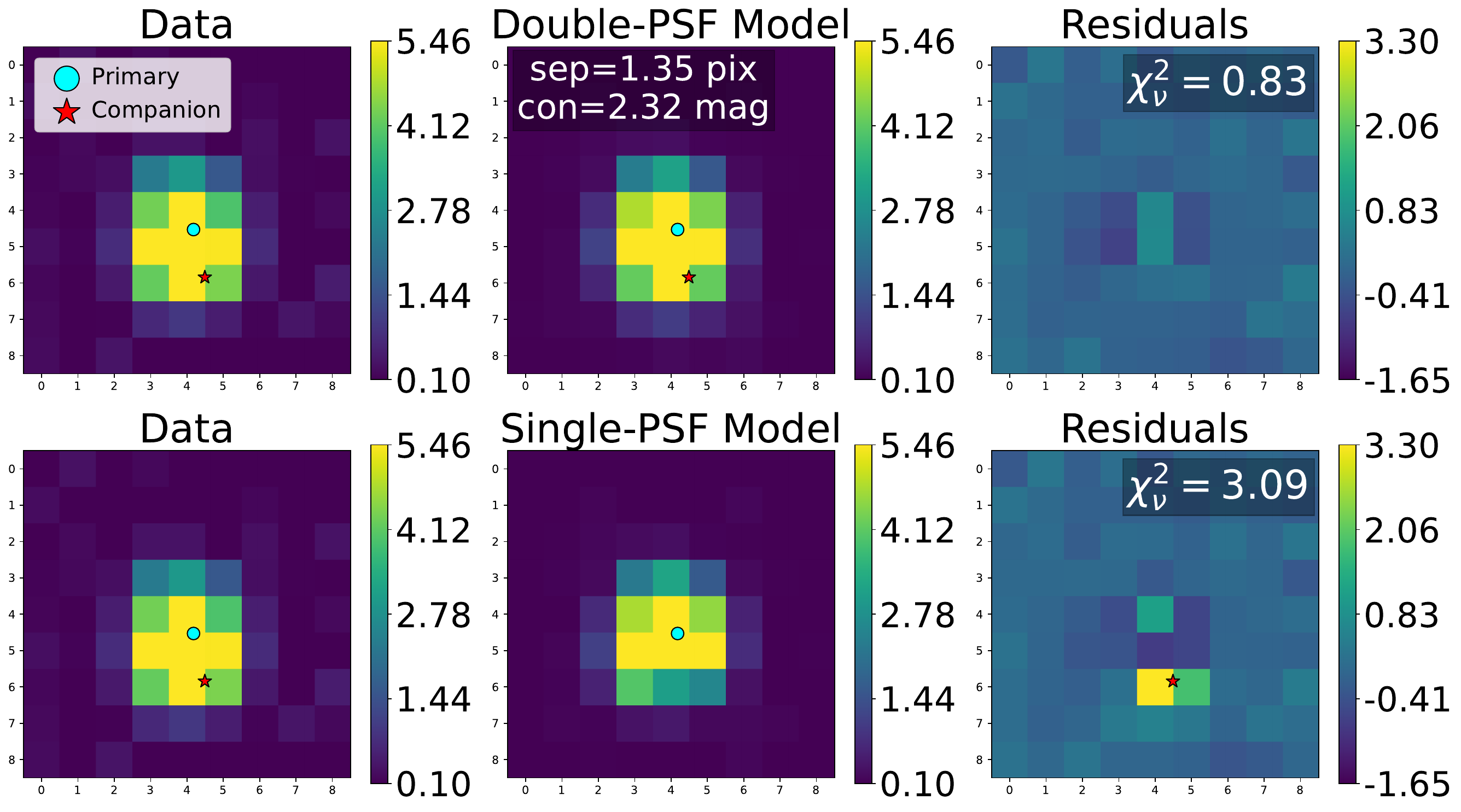}
    \caption{CFBDSIR 1458}
    \label{fig:cfbdsir1458_fits}
\end{subfigure}
\hfill
\begin{subfigure}[t]{0.47\textwidth}
    \includegraphics[width=\textwidth]{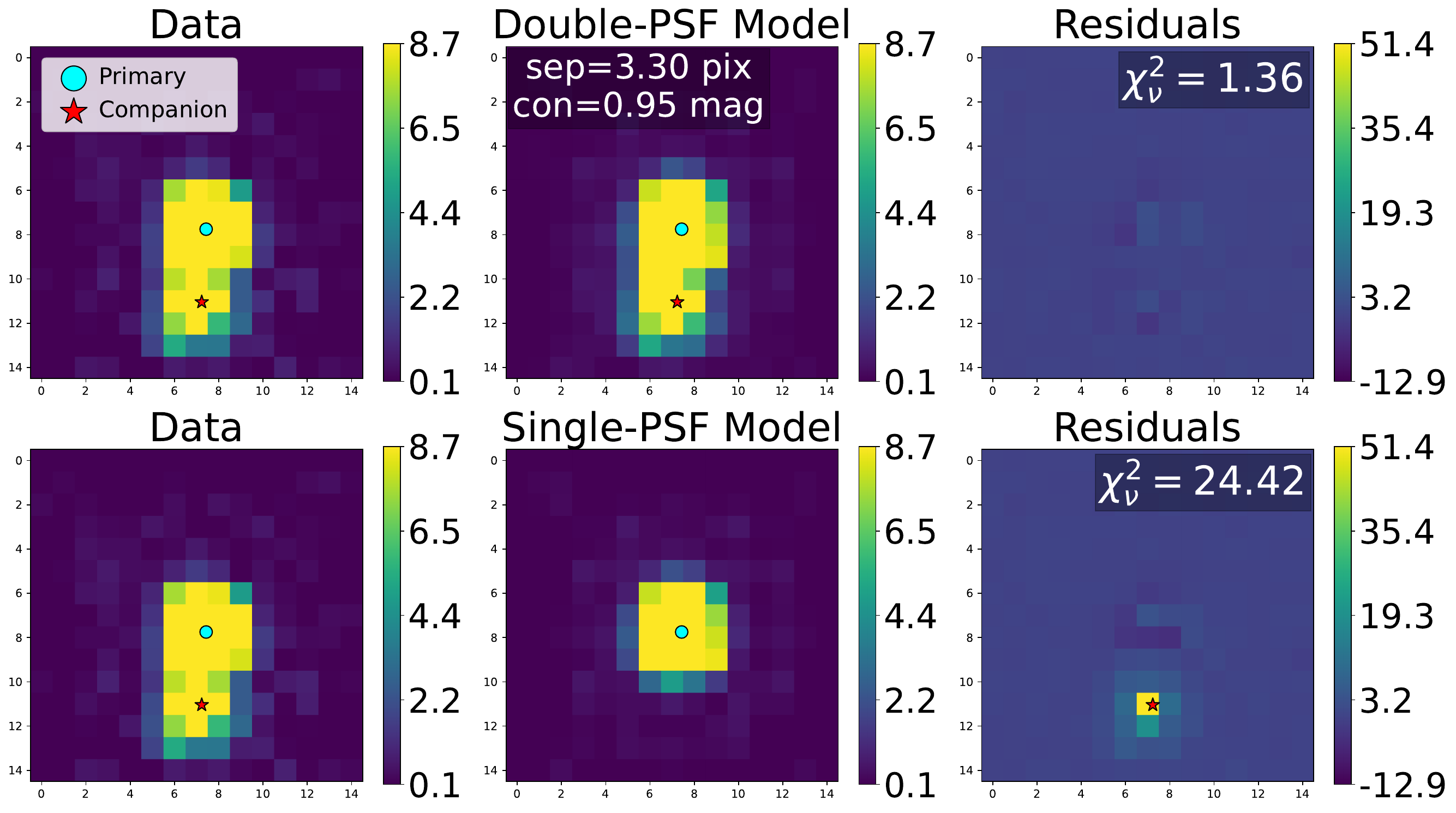}
    \caption{WISE 0458}
    \label{fig:wise0458_fits}
\end{subfigure}

\caption{PSF fitting to four example binary systems. The top row of each panel is a double-PSF fit and the bottom row is a single-PSF fit. The first column displays the data, the second displays the models, and the third displays the residuals. The reduced chi-squared for both fits is displayed in the Residuals plots. The cyan circle and red star are plotted at the locations of the primary and companion, respectively, from the double-PSF fit. Parameters from the latter three fits correspond to false positive and completeness probabilities that are indicative of recovered companions.}
\label{fig:fits_all}
\end{figure*}

\setcounter{table}{2}
\begin{deluxetable}{ccccccccc}
\tablecolumns{4}
\tablecaption{Recovered parameters of four example binaries.
\label{table:recovered_params}}
\tablehead{
  \colhead{\shortstack{Parameter\\Type}} &
  \colhead{Object} &
  \colhead{Filter} &
  \colhead{\shortstack{$\Delta x$ \\ (pix)}} &
  \colhead{\shortstack{$\Delta y$ \\ (pix)}} &
  \colhead{\shortstack{Magnitude \\ (Vega mag)}} &
  \colhead{\shortstack{PA \\ ($^{\circ}$)}} &
  \colhead{\shortstack{Separation \\ (pix)}} &
  \colhead{\shortstack{Contrast \\ (Vega mag)}} 
}
\startdata
\multirow{3}{*}{\centering \shortstack{Best Fit (max\\log-likelihood)}} & WISE 0146 & F127M & -0.36 & -1.05 & 23.22 & 228.23 & 0.81 & 0.99 \\
& CFBDSIR 1458 & F125W & 0.18 & -0.47 & 24.54 & 76.67 & 1.35 & 2.32 \\
& WISE 0458 & F160W & 0.43 & -0.25 & 23.56 & 93.46 & 3.30 & 0.95 \\
\hline
\multirow{4}{*}{\centering \shortstack{Posterior\\Distribution\\(Median $\pm1\sigma$)}} &WISE 0336&F125W& $-0.43^{+0.03}_{-0.03}$ & $0.36^{+0.05}_{-0.03}$ & $26.12^{+0.05}_{-0.05}$ & $199.37^{+77.15}_{-12.20}$ & $1.69^{+1.18}_{-0.51}$ & $2.44^{+0.76}_{-0.61}$ \\
&WISE 0146&F127M& $-0.36^{+0.04}_{-0.02}$ &  $-1.07^{+0.04}_{-0.04}$ & $23.22^{+0.003}_{-0.003}$ &  $228.18^{+9.88}_{-10.29}$ & $0.82^{+0.07}_{-0.05}$ &  $1.02^{+0.27}_{-0.29}$ \\
&CFBDSIR 1458&F125W&$0.12^{+0.01}_{-0.01}$ & $-0.47^{+0.01}_{-0.01}$ & $24.99^{+0.11}_{-0.05}$ & $78.05^{+4.35}_{-4.79}$ & $1.34^{+0.06}_{-0.12}$ & $2.27^{+0.11}_{-0.13}$ \\
&WISE 0458&F160W&$0.43^{+0.01}_{-0.25}$ & $-0.26^{+0.04}_{-0.25}$ & $23.56^{+0.001}_{-0.001}$& $93.79^{+3.60}_{-0.39}$ & $3.31^{+0.26}_{-0.01}$ & $0.96^{+0.01}_{-0.01}$
\enddata
\tablecomments{Best fit parameters of WISE 0336 omitted because S/N is outside the range characterized for this survey.}
\vspace{-3mm}
\end{deluxetable}

\subsection{WISE J014656.66+423410.0}
WISE J014656.66+423410.0 \citep{2015ApJ...803..102D}, hereafter WISE 0146, is a close and faint T9+Y0 system. From \cite{2015ApJ...803..102D}, imaging with Keck from 2013-10-22 places the separation at $92.9 \pm 4.0$ mas and imaging from 2012-09-07 and 2012-10-8 places the contrast in the \textit{J} band (closest analog to F127M) being $0.98 \pm 0.03$ and $1.07 \pm 0.08$ magnitudes respectively. This corresponds to a separation in WFC3/IR of $0.71 \pm 0.03$ pixels. We chose this system to test recovery of a close separation and low contrast companion. 

Applying both our technique and a single-PSF fit, we obtain results shown in Fig. \ref{fig:wise0146_fits}. The S/N in F127M for this object calculated using eq. \ref{eq:SNR} is $S/N \approx 35$. We plot the best fit separation and contrast (see table \ref{table:recovered_params}) on the F127M and $S/N=35$ completeness probability heatmap (see Fig. \ref{fig:wise0146_plot}). The best fit values fall in a region of parameter space where the detection has a $<1\%$ chance of being a false positive and where we confidently recover companion parameters $\sim45$\% of the time. Given moderate confidence in a detection and a posterior separation consistent within $2\sigma$ and contrast within $1\sigma$ of observed parameters from a close-in-time observation, this fit demonstrates sensitivity to a companion close to the limit for this technique. The best fit separation along with updated parallax measurements \citep{2021ApJS..253....7K, 2025ApJ...984...74B} place the projected separation at $2.06 \pm 0.17$ AU in this observation. Therefore, we display the ability to detect companions down to $0.96\lambda$/D at 1 mag in contrast for low S/N values in F127M. 


\renewcommand{\thefigure}{7}
\begin{figure*}[t!]
\centering
\begin{subfigure}[t]{0.49\textwidth}
    \includegraphics[width=\textwidth]{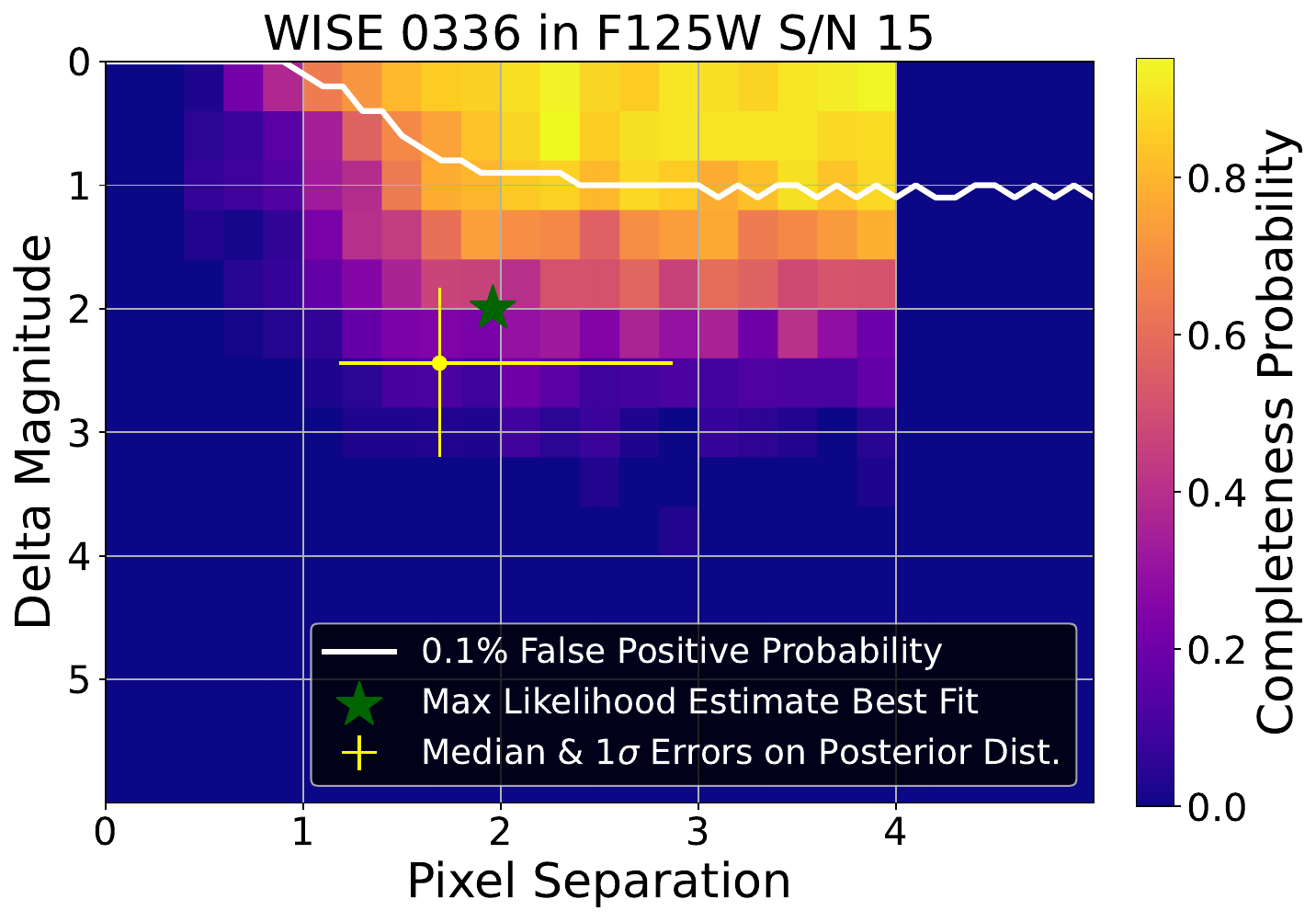}
    \caption{WISE 0336}
    \label{fig:wise0336_plot}
\end{subfigure}
\hfill
\begin{subfigure}[t]{0.49\textwidth}
    \includegraphics[width=\textwidth]{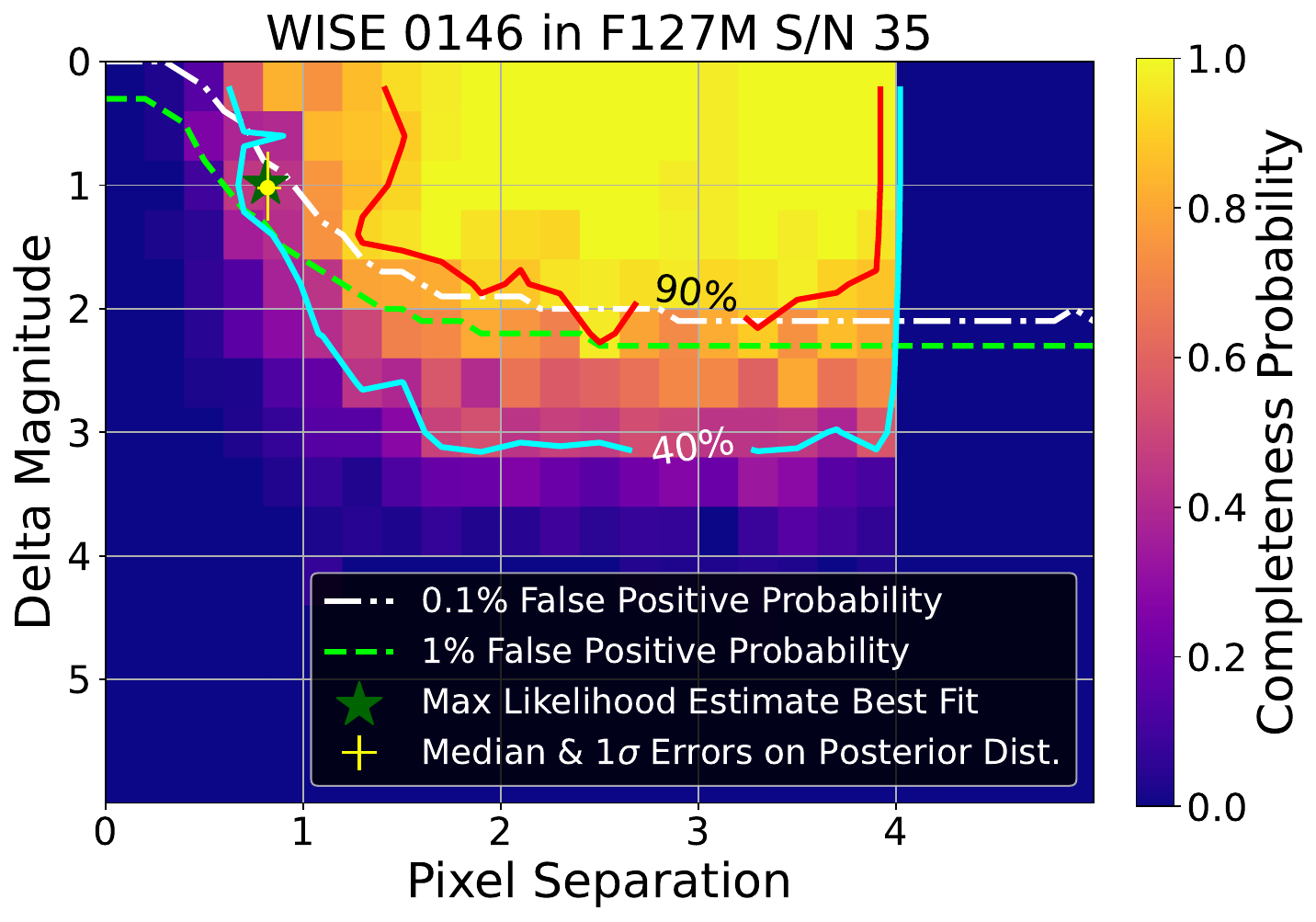}
    \caption{WISE 0146}
    \label{fig:wise0146_plot}
\end{subfigure}
\begin{subfigure}[t]{0.49\textwidth}
    \includegraphics[width=\textwidth]{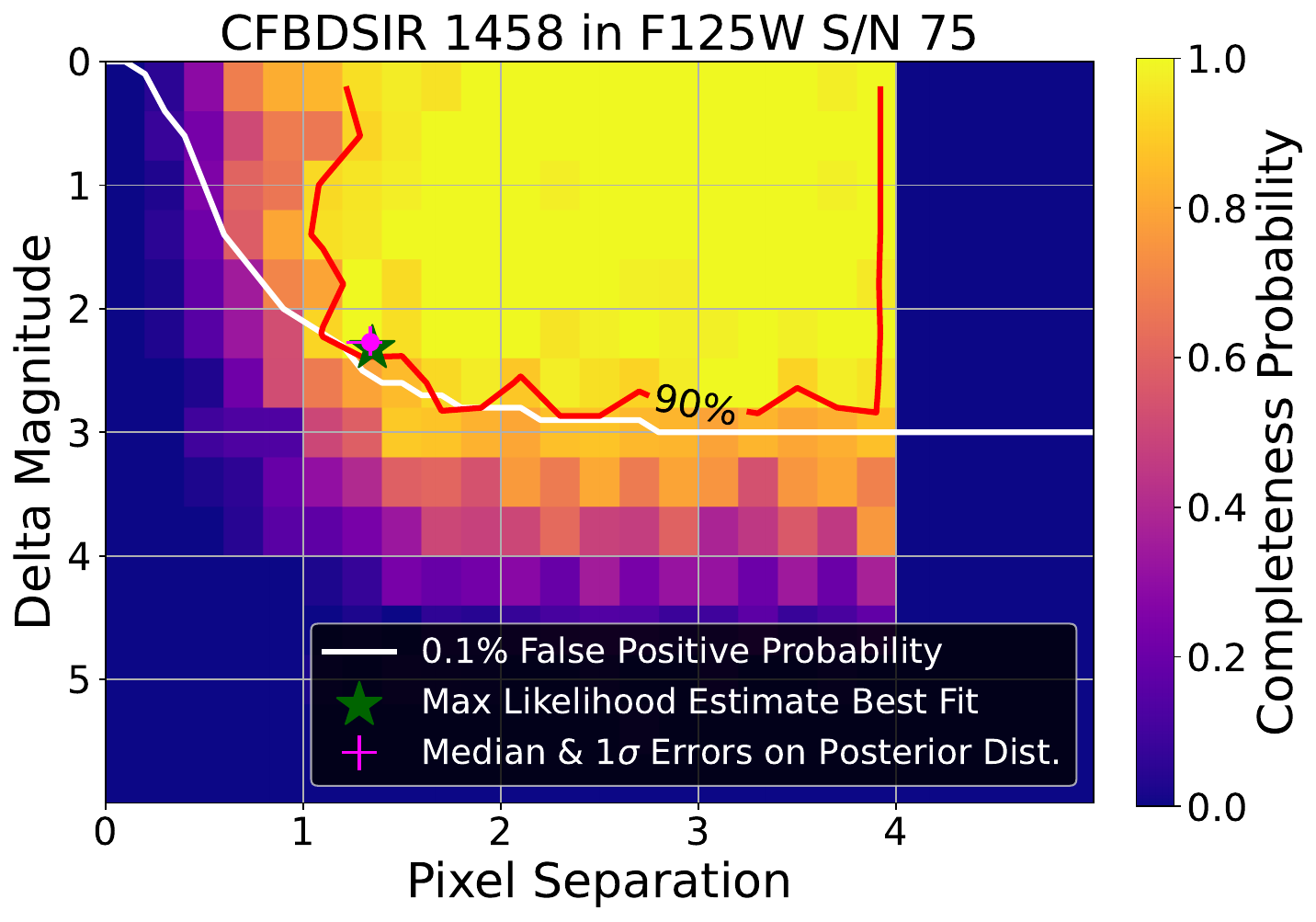}
    \caption{CFBDSIR 1458}
    \label{fig:cfbdsir1458_plot}
\end{subfigure}
\hfill
\begin{subfigure}[t]{0.49\textwidth}
    \includegraphics[width=\textwidth]{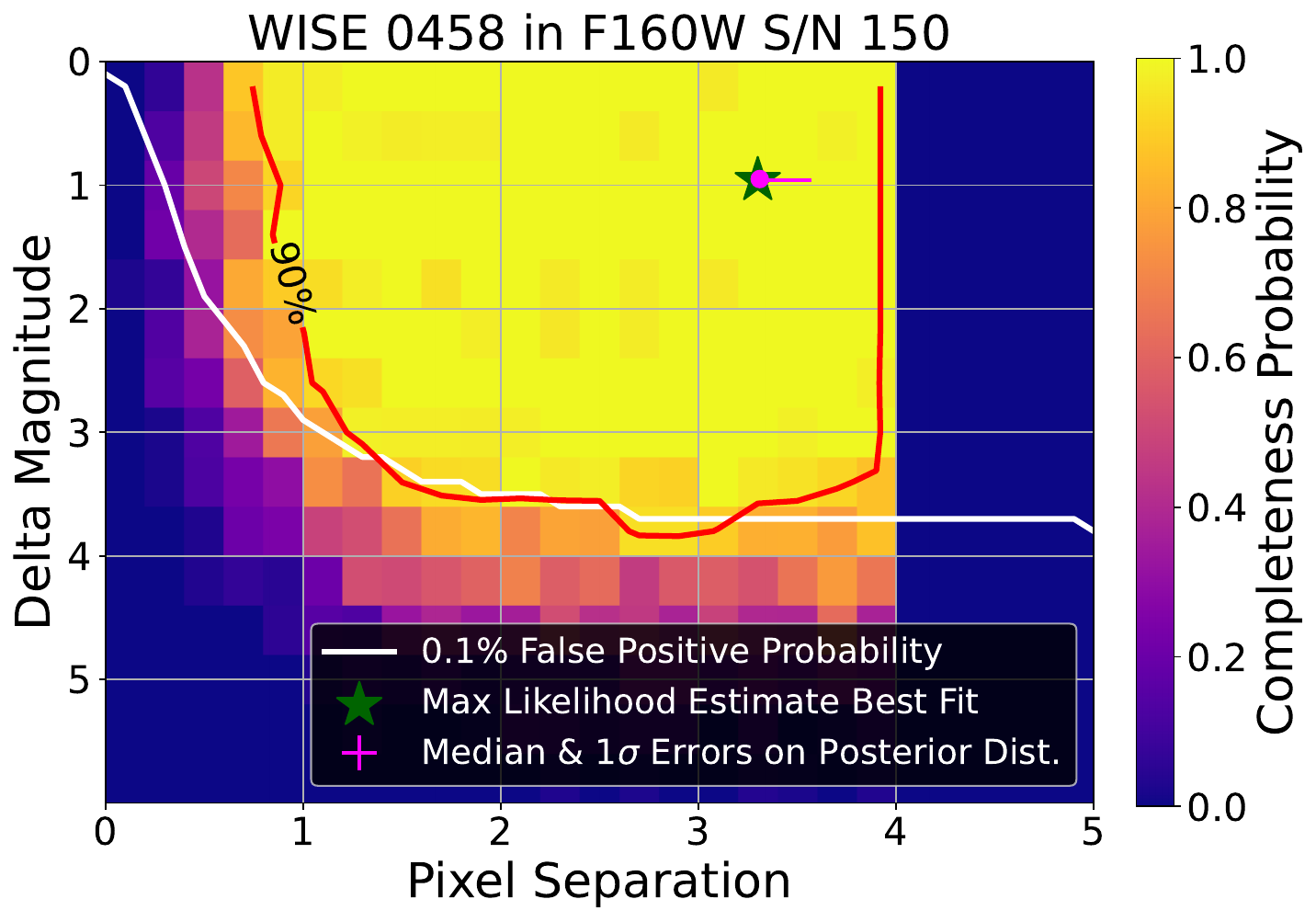}
    \caption{WISE 0458}
    \label{fig:wise0458_plot}
\end{subfigure}

\caption{Completeness probability maps in separation and contrast space for the four example binary systems in the filter of observation and S/N of the object. Overplotted are the 0.1\% false positive probability lines in white and 90\% completeness probability lines in red. Due to discontinuities in the S/N 15 completeness lines (see Sec. \ref{subsec:completecurves}), the curve has been omitted for Panel a).} Panel b) additionally displays the 40\% completeness probability line in aqua and the 1\% false positive probability line dotted in green. The recovered best fit separation and contrast of each are plotted as the green star along with the median of the posterior distributions in yellow in panels a) and b) and magenta in panels c) and d).
\label{fig:fits_all}
\end{figure*}

\subsection{CFBDSIR J145829+101343}
CFBDSIR J145829+101343 \citep{2011ApJ...740..108L}, hereafter CFBDSIR 1458, is a T9+Y0 system with a separation of $127.2 \pm 1.4$ mas and contrast in the \textit{J} band of $2.02 \pm 0.07$ magnitudes in observations with Keck taken on 2012-04-13 \citep{2012ApJ...758...57L}. This corresponds to a separation in WFC3/IR of $0.98 \pm 0.01$ pixels. We chose this system to test recovery of a moderate separation and contrast companion. 

Applying both our technique and a single-PSF fit, we obtain results shown in Fig. \ref{fig:cfbdsir1458_fits}. The S/N in F125W for this object calculated using eq. \ref{eq:SNR} is $S/N \approx 75$. We plot best fit separation and contrast (see table \ref{table:recovered_params}) on the F125W and $S/N=75$ completeness probability heatmap (see Fig. \ref{fig:cfbdsir1458_plot}). The best fit values fall in a region of parameter space where the detection has a $<0.1\%$ chance of being a false positive and where we confidently recover companion parameters $>90$\% of the time. Given this confidence in the detection and the posterior median separation and contrast consistent within $2\sigma$ of a close-in-time observation, we display likely recovery of a companion. The best fit separation along with parallax measurements from \cite{2011ApJ...740..108L} place the projected separation at $4.02 \pm 0.50$ AU in this observation. Therefore, our technique can resolve down to 1.6 $\lambda$/D at 2.25 mag in contrast in F125W for a moderate S/N source. 

\subsection{WISEPA J045853.89+643452.9}
WISEPA J045853.89+643452.9 \citep{2011ApJ...726...30M}, hereafter WISE 0458, is a T8+T9.5 system with observed companion parameters of $455.1 \pm 4.2$ mas in separation and $1.02 \pm 0.01$ in the \textit{H} band taken with Keck NIRC2 on 2011-08-29 \citep{2012ApJ...745...26B}. This corresponds to a separation in WFC3/IR of $3.50 \pm 0.03$ pixels. We chose this system to test recovery of wide and low contrast systems that are also identified via more classical analysis methods. 

Applying both our technique and a single-PSF fit, we obtain results shown in Fig. \ref{fig:wise0458_fits}. The S/N in F160W for this object calculated using eq. \ref{eq:SNR} is $S/N \approx 177$. We plot the best-fit separation and contrast (see table \ref{table:recovered_params}) on the F160W $S/N=150$ completeness probability heatmap (see Fig. \ref{fig:wise0458_plot}). The best-fit values comfortably fall in a region of parameter space where the detection has a $<0.1\%$ chance of being a false positive and where we confidently recover companion parameters $>90$\% of the time. Given strong confidence in a detection and a posterior median separation consistent within $1\sigma$ and contrast within $2\sigma$ of observed parameters from a close-in-time observation, we display comfortable recovery of a wide companion. The best fit separation along with parallax measurements from \cite{2021ApJS..253....7K} place the projected separation at $4.03 \pm 0.20$ AU in this observation. Therefore, our technique can easily recover companions at 3.1 $\lambda$/D at 1 mag in contrast in F160W for a high S/N source. 

\section{Discussion}\label{sec:discussion}
\subsection{Sensitivity Comparisons}\label{subsec:comparisons}
 Our results display the ability of the double-PSF fitting algorithm to resolve companions in the subpixel regime across a range of filters. Even in the cases where subpixel sensitivity to companions is not possible, we have significantly improved our sensitivity over previous surveys in HST near-IR imaging.

These surveys \citep[e.g.,][]{2014AJ....148..129A,10.1093/mnras/stad2870} utilize PSF subtraction and fitting with sensitivity defined by artificial star tests and Monte Carlo simulations, respectively. The improvement in our technique arises from the ability to model and fit both PSFs simultaneously. Fitting and subtracting the primary will produce high residuals and remove PSF structure of close companions, limiting the sensitivity of the other surveys. Furthermore, we use position dependent PSF models empirically derived from rich star fields which produce better fits uniformly across the detector than empirical models derived from background sources or the targets themselves.
\renewcommand{\thefigure}{8}
\begin{figure}[h]
    \centering
    \includegraphics[width=\columnwidth]{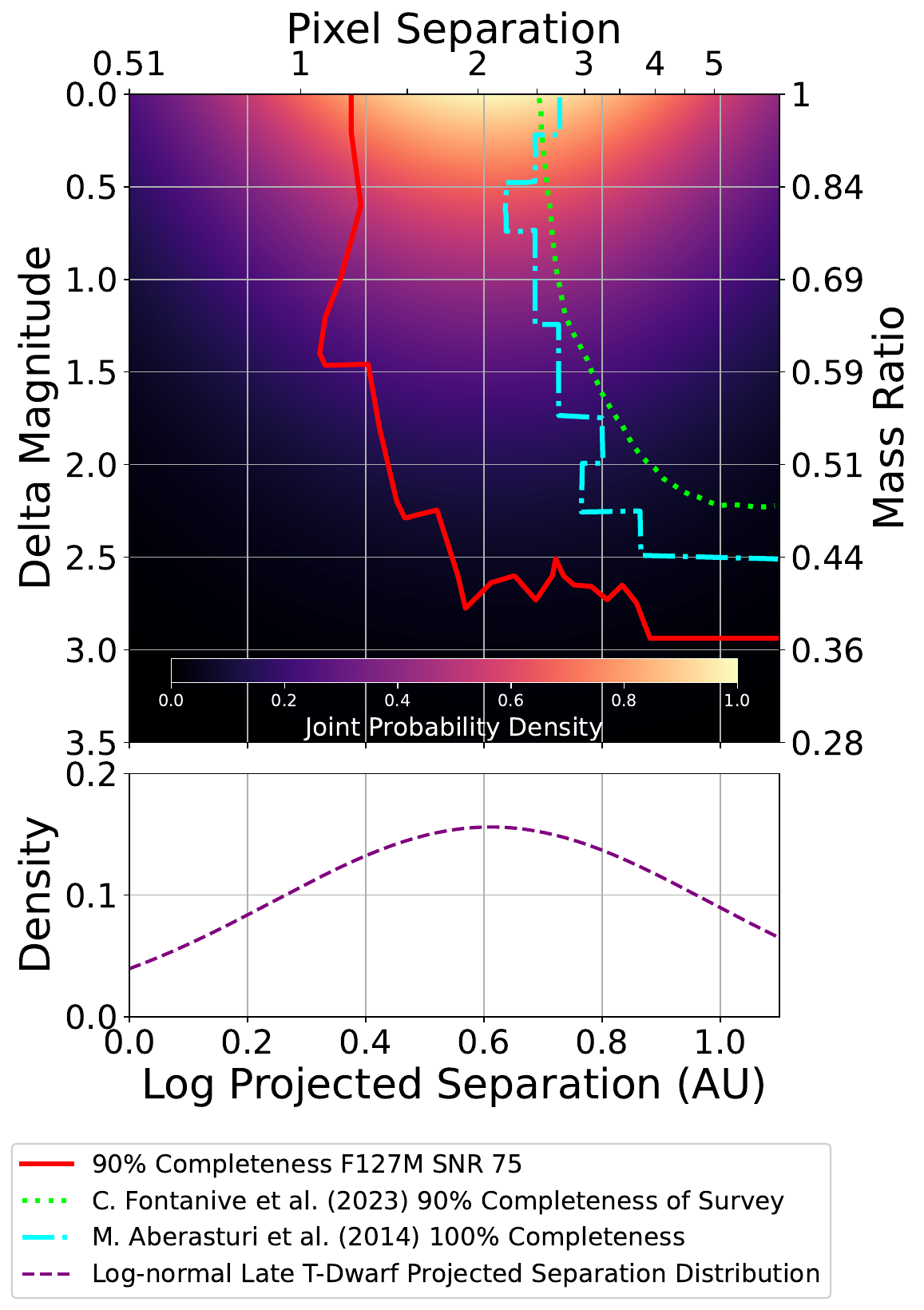} 
    \caption{Completeness curves in F127M from our sensitivity analysis (in red) and previous surveys (in cyan and green) plotted in separation and contrast space compared against the joint probability heatmap of the log-normal projected separation distribution of late T-dwarfs (Fig. 22 of \cite{10.1093/mnras/sty1682}) convolved with the mass ratio power law \citep{2006AJ....132..891R}. The T-dwarf separation distribution PDF is shown in the bottom marginal plot. Separation is plotted in both pixels and AU (projected to a distance of 15 pc). Mass ratio is shown on the right y-axis, derived from ATMO 2020 \citep{2020A&A...637A..38P} evolutionary models assuming an age of $\sim 1.2$ Gyr and primary mass of $0.041 M_{\odot}$.}
    \label{fig:completeness_comp}
\end{figure}
Both surveys provide completeness curves for F127M which are plotted and compared with our F127M curve with a S/N of 75 in Fig. \ref{fig:completeness_comp}---a median representation of sensitivity in the filter. These curves are compared to a log-normal projected separation distribution of late T dwarfs from Fig. 22 of \cite{10.1093/mnras/sty1682} assuming a distance of 15 parsecs. We also convolved a separation and a mass ratio distribution (assuming a primary of 0.041 $M_{\odot}$, shown as the heatmap. We are 2.5$\times$ more sensitive in separation over contrasts of 0-3 magnitudes in F127M compared to previous surveys with similar methods. When comparing completeness bounds, we show an improvement in our ability to recover companions from $\sim 5$ AU to $\sim 2.25$ AU, assuming 15 pc. This advancement comfortably encompasses the peak of the inferred late T dwarf separation distribution as well as the convolved separation-mass ratio distribution, covering a previously unsampled range of the parameter space in HST/WFC3 imaging and opening up the possibility of detecting closer companions to conceivably better constrain the separation distribution. 

We can compare this technique to analyses on other high angular resolution imaging telescopes. In \cite{2023ApJ...948...92D}, adaptive optics imaging of the Y-dwarf WISE 1828+2650 is conducted with Keck/NIRC2. In the \textit{H} band, sensitivity to companions at an angular separation of 71 mas (0.55 pixels in WFC3/IR) is achieved at 1.5 magnitudes contrast which corresponds to an angular resolution of $2.15\lambda/D$. The HST filter analog for the \textit{H} band is F160W \citep{WFC3IHB_V17.0_2024} where we achieve sub-diffraction limited imaging. Although we do not achieve similar separations due to Keck's 4$\times$ larger aperture, we are still competitive with Keck as we can recover companions down to 0.1" in the high S/N case. This demonstrates that smaller space-based telescopes have such a stable PSF that they can routinely detect companions at closer separations in $\lambda/D$ space than large ground based telescopes with AO, for very faint targets.

Comparing our technique's ability against other methods of resolving close point sources, the ability and practicality of our technique sets itself apart. Kernel Phase Interferometry (KPI) is a technique that can achieve angular separations of $\sim0.5\lambda/D$ \citep{2010ApJ...724..464M,2022AJ....164..244F,2023AJ....165..130F,2013ApJ...767..110P, 2023PASP..135a4502K}.
Specifically, \cite{2010ApJ...724..464M} tests KPI on HST/NICMOS imaging of GJ 184, recovering a low mass companion to an M dwarf primary. The binary has a combined \textit{H} band apparent magnitude of $8.248 \pm 0.030$ magnitudes and contrast of $1.835 \pm 0.006$ magnitudes \citep{2009ApJ...695.1183M}. 
Given a separation of $\sim88$ mas (0.68 pixels in WFC3/IR) and contrast of $1.84$ magnitudes, we could recover the companion with $\sim80\%$ confidence at $0.64\lambda/D$ in the \textit{H} band for high S/N values. Another method is aperture masking interferometry (AMI), which can resolve down to $0.3 \lambda/D$ with direct imaging in the K band \citep{2024A&A...682A.101S, 2025ApJ...983L..25R, 2024ApJ...963L...2S, 2019JATIS...5a8001S, 2020ApJ...889..175R}. Sparse AMI can achieve even better sensitivity with a $\lambda/D$ of $0.0046$ in the $L^{\prime}$ band on the Very Large Telescope \citep{2011A&A...532A..72L}. 

While our technique applied in WFC3/IR does not attain as low angular separations as KPI or AMI, these methods require calibration sources and are performed on a source by source basis, while our technique achieves relatively comparable resolutions and can be applied to many sources imaged simultaneously across the entire wide-field camera (e.g., in a star cluster). For certain use cases, our technique may present a more preferable choice over KPI and AMI. Furthermore, not only can AMI not be performed on HST, but KPI has not been used to analyze WFC3/IR data due to the undersampled nature of the detector. Therefore, we achieve the highest angular resolution to date in WFC3/IR imaging which highlights the significant potential of this technique given the extensive archival dataset accumulated over 15 years of WFC3/IR operations.

\subsection{Limitations and Observations}\label{subsec:limitations}
We recognize that our sensitivity analysis is limited due to the poor sampling of WFC3/IR which causes much of the PSF of an object to land within one pixel. Losing PSF detail impacts the ability of our technique to fit closely separated PSFs since there is less granularity in the data when much of the PSF structure falls under one data point. This overly impacts low S/N objects, restricting our ability in contrast space. While resampling methods such as drizzling \citep{2002PASP..114..144F} images can traditionally fix this issue, resampling changes the PSF shape leading to poor fits with empirical PSF models \citep{2016wfc..rept...12A}. Another option to achieve similar sensitivity is to fit all dither positions simultaneously, achieving a similar S/N as a drizzled image on one object without resampling. 

We observe an interesting phenomenon with the false positive heatmaps which are particularly highlighted by the 0.1\% curves. For the filters F105W, F110W, and F127M and S/N values of 75, 150, and 300, the false positive lines display relatively worse performance than the same S/N curves in other filters, being shifted up $\sim1$ magnitude with respect to the background limit contrast. The most evident feature is that the S/N 35 and 75 curves display the same performance in these three filters. This phenomenon also occurs in Fig. 3 of \cite{2022ApJ...925..112D} where S/N 50 and 70 curves in F435W, F555W, F775W, and F850LP. This is likely due to the location of the first diffraction minimum for these passbands. As S/N increases, the dominant noise shifts from detector and sky noise to photon noise which produces less spread out residuals with more noise occurring in the PSF core inside the first diffraction minimum. In wavelengths where the diffraction limit is $\sim 1$ pixel, the buildup of photon noise residuals at this location leads to a likely over-convergence of false positive fits. This yields more fits to single PSFs converging at lower contrasts and closer separations, forcing the false positive curves lower in contrast and wider in separation. It is important to consider that the false positive and completeness percentages we choose to display are conservative assumptions and not necessarily the values that must be used. Different values for the bounds can be chosen depending on sample sizes and individual use cases. 
\subsection{Our Survey}\label{subsec:survey}
This analysis is one part of a larger effort (The Hubble Ultracool Multiplicity Survey) to search for faint and close companions to BDs. With the abilities of our technique set forth in this paper, we will conduct one of the largest surveys of field BDs to date and robustly process images of a large number of BDs to better constrain multiplicity statistics, including companion frequency, separation distribution, and mass ratio distribution. Importantly, this survey will be conducted on HST archival data which requires no additional observing time and possesses a significant range of previously unsampled separation space which is newly accessible with our technique. With a plethora of BD images in the HST archive spanning a variety of instruments, the sample size can be further expanded by characterizing the sensitivity across other detectors using the same method described in this paper. With WFC3/IR and ACS/WFC \citep{2022ApJ...941..161D} characterized, we are currently working on WFPC2/PC and WFC3/UVIS. For other targeted science cases, we may characterize ACS/HRC in the future which contains less archival BD data than other cameras but has samples present in star-forming regions and young clusters. We will apply the fitting technique to data in these instruments and utilize the sensitivity analyses to vet candidate detections. This is the first in a series of papers that will produce the most robust multiplicity demographics for field BDs to date. 
\subsection{Other Applications}\label{subsec:futureapplication}
The technique used in this paper is versatile in its applications. The main dependency that the fitting operates on is having the ability to produce empirical PSF models for stable telescopes. Therefore, this technique can be applied to images taken with any space-based telescope in any camera and filter. One can even make contemporaneous empirical PSF models as shown in \cite{2023ApJ...947L..30C,2023ApJ...948...92D}. 

One particular strength of our technique can be shown in star forming regions (SFR). Since our algorithm can effectively recover companions near the diffraction limit, we are able to probe these regions and accurately define statistics like multiplicity frequency in young objects \citep{2022ApJ...925..112D}. Our method is especially useful because many sources in the SFR will fall within one image due to the field of view, allowing for fits to more sources with less observations than imaging sources individually as is needed for techniques like KPI or AMI. Furthermore, this technique can be used to deblend confused PSFs and recover astrometry and photometry of objects which are in crowded stellar fields---too closely separated for other methods. This ability proves useful to understand the composition of dense SFRs and globular clusters and extract information necessary to explore the initial mass function. Since our technique is applicable to any situations with two PSFs, it can also be applied to the search for exoplanets around their host stars, provided that the contrast is within recoverable limits. 

Looking forward to the next generation of telescopes, this technique can be applied similarly to the WFI on the Nancy Grace Roman Telescope. The instrument will have nearly 200$\times$ the field-of-view of WFC3/IR and will be better sampled with a smaller plate scale. Such benefits have already proven valuable: JWST's larger aperture and deeper sensitivity has allowed for higher angular resolution \citep{2023ApJ...947L..30C, 2025ApJ...990L..63D}. Furthermore, the PSF of Roman should be very stable, which will make it easier to create PSF libraries populated with robust empirical models. The imminent Galactic Plane Survey \citep{2025arXiv251107494G} will provide imaging of many BDs, both in the field and in star forming regions, at improved resolution where this technique can be applied with greater effectiveness to search for companions. Better resolution will also allow for longer orbit monitoring of binaries since more orbits can be tracked along the inward movement of a companion.

\section{Conclusion}\label{sec:conclusion}
We characterized the sensitivity for our double-PSF fitting technique in WFC3/IR on HST. In summary: 
\begin{enumerate}
    \item We utilized empirical, position dependent PSFs to create synthetic single and double point sources and applied our technique to the synthetic data to define false positive and completeness probabilities in separation and contrast parameter space across 5 S/N values in 8 filters.
    \item We found that in all filters analyzed, particularly for high S/N values, the technique can recover companions at or below the diffraction limit ($\sim 0.10^{\prime\prime} - 0.17 ^{\prime\prime}$ or 0.77-1.30 pixels for F098M - F160W). This corresponds to a 2.5$\times$ improvement on previous surveys that analyzed the same data \citep{10.1093/mnras/stad2870, 2014AJ....148..129A}. We achieve the highest angular resolution to date in WFC3/IR imaging for any method. 
    \item We apply the technique to four known binaries: WISE J033605.05-014350.4, WISE J014656.66+423410.0, CFBDSIR J145829+10134, and WISEPA J045853.89+643452.9. We detect companions for the latter three at the expected separations and contrasts of each system. We show the detection of a companion to WISE 0146 down to $0.96\lambda$/D at 1 mag in contrast for F127M. This demonstrates our ability to recover companions at sub-diffraction limit scales, displaying improvements over previous analyses.
\end{enumerate}

\section{Acknowledgements}
DBG and MDF acknowledge support from HST-AR-17561 (co-PIs: Bardalez Gagliuffi, De Furio). KM acknowledges support from the National Science Foundation (NSF) Research Program for Undergraduates (REU) grant AST-2244278 (PI:Jogee). M.D.F. is supported by an NSF Astronomy and Astrophysics Postdoctoral Fellowship under award AST-2303911. 

This research is based on observations made with the NASA/ESA Hubble Space Telescope obtained from the Space Telescope Science Institute, which is operated by the Association of Universities for Research in Astronomy, Inc., under NASA contract NAS 5–26555. These observations are associated with programs GO-12970, SNAP-12873, GO-12504, and GO-17466.


\bibliography{sample701}{}
\bibliographystyle{aasjournalv7}



\end{document}